\documentclass{aastex7}

\usepackage{longtable} 
\usepackage{caption} 
\usepackage{adjustbox}
\usepackage{tabularx}
\usepackage{multirow}
\usepackage{natbib}
\usepackage{array}

\begin{document}

\title{The Milky Way-like radial structure in nearby galaxies}

\author{Yuze Zhao}
\affiliation{South-Western Institute for Astronomy Research, Yunnan University, Kunming, Yunnan 650091, China}
\email{}  

\author[0000-0001-5258-1466]{Jianhui Lian}
\affiliation{South-Western Institute for Astronomy Research, Yunnan University, Kunming, Yunnan 650091, China}
\affiliation{Yunnan Key Laboratory of Survey Science, Yunnan University, Kunming, Yunnan 650500, China}
\email{jianhui.lian@ynu.edu.cn} 

\author{Zheng Zheng}
\affiliation{National Astronomical Observatories, Chinese Academy of Sciences, Beijing 100101, China}
\affiliation{Key Laboratory of Radio Astronomy and Technology, Chinese Academy of Sciences, Beijing, 100101, China}
\email{} 

\author[0000-0001-6405-9675]{Cuihuan Li}
\affiliation{National Astronomical Observatories, Chinese Academy of Sciences, Beijing 100101, China}
\email{} 

\author[0000-0002-0846-7591]{Jie Song}
\affiliation{Department of Astronomy, University of Science and Technology of China, Hefei 230026, China}
\affiliation{School of Astronomy and Space Science, University of Science and Technology of China, Hefei 230026, China}
\affiliation{Institute of Deep Space Sciences, Deep Space Exploration Laboratory, Hefei 230026, China}
\email{} 

\author[0009-0005-8762-0871]{Tao Wang}
\affiliation{South-Western Institute for Astronomy Research, Yunnan University, Kunming, Yunnan 650091, China}
\email{} 

\author[0009-0002-9360-2280]{Zhenkang Tang}
\affiliation{South-Western Institute for Astronomy Research, Yunnan University, Kunming, Yunnan 650091, China}
\email{} 

\author{Dezi Liu}
\affiliation{South-Western Institute for Astronomy Research, Yunnan University, Kunming, Yunnan 650091, China}
\affiliation{Yunnan Key Laboratory of Survey Science, Yunnan University, Kunming, Yunnan 650500, China}
\email{} 

\author{Xiangkun Liu}
\affiliation{South-Western Institute for Astronomy Research, Yunnan University, Kunming, Yunnan 650091, China}
\affiliation{Yunnan Key Laboratory of Survey Science, Yunnan University, Kunming, Yunnan 650500, China}
\email{}

\author{Yuan Fang}
\affiliation{South-Western Institute for Astronomy Research, Yunnan University, Kunming, Yunnan 650091, China}
\affiliation{Yunnan Key Laboratory of Survey Science, Yunnan University, Kunming, Yunnan 650500, China}
\email{}

\author{Guowang Du}
\affiliation{South-Western Institute for Astronomy Research, Yunnan University, Kunming, Yunnan 650091, China}
\affiliation{Yunnan Key Laboratory of Survey Science, Yunnan University, Kunming, Yunnan 650500, China}
\email{}

\author{Jinghua Zhang}
\affiliation{South-Western Institute for Astronomy Research, Yunnan University, Kunming, Yunnan 650091, China}
\affiliation{Yunnan Key Laboratory of Survey Science, Yunnan University, Kunming, Yunnan 650500, China}
\email{}

\author{Xufeng Zhu}
\affiliation{South-Western Institute for Astronomy Research, Yunnan University, Kunming, Yunnan 650091, China}
\affiliation{Yunnan Key Laboratory of Survey Science, Yunnan University, Kunming, Yunnan 650500, China}
\email{}

\author{Xiaowei Liu}
\affiliation{South-Western Institute for Astronomy Research, Yunnan University, Kunming, Yunnan 650091, China}
\affiliation{Yunnan Key Laboratory of Survey Science, Yunnan University, Kunming, Yunnan 650500, China}
\email{}

\author{Xinlei Chen}
\affiliation{South-Western Institute for Astronomy Research, Yunnan University, Kunming, Yunnan 650091, China}
\email{}

\author{Yu Pan}
\affiliation{South-Western Institute for Astronomy Research, Yunnan University, Kunming, Yunnan 650091, China}
\email{}

\author{Xingzhu Zou}
\affiliation{South-Western Institute for Astronomy Research, Yunnan University, Kunming, Yunnan 650091, China}
\email{}

\author{Ye-Hao Cheng}
\affiliation{South-Western Institute for Astronomy Research, Yunnan University, Kunming, Yunnan 650091, China}
\email{}

\author{Shuai Xie}
\affiliation{South-Western Institute for Astronomy Research, Yunnan University, Kunming, Yunnan 650091, China}
\email{}

%\collaboration{all}{The Terra Mater collaboration}

\begin{abstract}

The Milky Way exhibits a distinctive plateau in its inner surface brightness profile around one effective radius, challenging the traditional view of a purely exponential disk. To explore the origin of this feature, we identify 65 nearby galaxies ($<50$ Mpc) with similar plateau structures and analyze their multi-band surface brightness and color profiles using SDSS, DESI, 2MASS, and Mephisto data. Of the 65 galaxies with Milky Way-like plateaus, 42 have bar classifications (25 barred, 17 non-barred). Barred galaxies show radial color variations more consistent with stellar population ageing, while non-barred galaxies align more with the dust reddening direction, suggesting dust attenuation contributes to plateau formation in non-barred systems. The broad range of color gradients among barred galaxies suggests bars influence plateaus through multiple pathways, including stellar population variations and dynamical stellar-mass redistribution. Environment may also contribute: both barred and non-barred galaxies tend to reside in denser environments than a comparison sample without plateaus. The low occurrence rate ($\sim$3.6\%) may indicate that plateau structures mark a relatively uncommon evolutionary stage. The Milky Way, with both strong stellar population dependence and a persistent stellar mass profile feature, may represent an intermediate case where both stellar population variations and dynamical processes associated with the bar shape its plateau.

\end{abstract}

%\keywords{\uat{Galaxies}{573} --- \uat{} {}}



\section{INTRODUCTION} 
 
The radial structure of a galaxy provides critical constraints on its assembly history, with exponential disks long recognized as a fundamental feature of galaxy structure \citep{1970_Freeman, 1980Fall}. Early studies of edge-on galaxies revealed deviations from pure exponential profiles in the form of disk truncations \citep{1979van_der_Kruit}, while subsequent deep imaging studies demonstrated a diversity of surface brightness profiles \citep{2002Pohlen, 2005Erwin, 2008Erwin}. Based on systematic analysis of radial surface brightness profiles, \citet{2006Pohlen} classified approximately 90 face-on nearby late-type spiral galaxies into three categories: 10\% Type I (single exponential), 60\% Type II (down-bending), and 30\% Type III (up-bending). Both Type II and Type III profiles are characterized by an outer break radius, beyond which the radial gradient becomes steeper (Type II) or shallower (Type III). 
Many studies have reported similarly high fractions of galaxies with broken exponential profiles (Type II and Type III), such as 88\% \citep{Guti2011}, 84\% \citep{Marino2016}, and 76\% \citep{2020Tang}, indicating that surface brightness profile breaks are a common structural feature of disk galaxies.
These fractions, however, are not constant but vary with environment \citep{environment} and redshift \citep{Xu_Yu_2024}. An intriguingly suppressed frequency of Type II disk is found in Perseus cluster, which is attributed to tidal effect. \citet{Zheng_2015} further showed, using a sample of 698 low-redshift disk galaxies, that most systems exhibit Type II profiles in bluer bands, with break radii typically located around $\rm r_{90}$ (the radius containing 90\% of the r-band flux). In contrast, surface brightness profiles in redder bands, as well as stellar mass surface density profiles, are generally consistent with a single exponential form. This behavior is consistent with previous studies showing that Type II disks exhibit ``U-shaped'' color profiles, and that the corresponding breaks in surface brightness profiles largely disappear in stellar mass surface density profiles \citep{2008Bakos}. 

Type I disks are often interpreted as the outcome of angular momentum conservation during disk formation or viscous redistribution processes \citep{1987van_der_Kruit, 2001Ferguson}. Type II breaks have been attributed to several physical mechanisms, including star formation thresholds \citep{1989Kennicutt, 2004Schaye, 2006Elmegreen, 2012Martin} and stellar radial migration \citep{2006Debattista, 2008Roskar}, as well as inside-out disk growth \citep{down-bending_inside-out}. Type III breaks are likely driven by environmental influences, most notably tidal disturbances that induce morphological asymmetry \citep{up_bending}. Despite this well-established phenomenological framework, the physical origin of deviations from exponential profiles is not completely understood.  

Galaxy bars likely play a key role in shaping the galactic disk structure. They can modify galaxy radial profiles through two main pathways. First, bars can induce dynamical redistribution of stars through angular momentum exchange, altering the stellar mass distribution. For example, simulations by \citet{2021thor} show that bar-driven secular evolution can transform an initially exponential disk into a composite profile consisting of an inner Type III and an outer Type II component. Second, bars can regulate star formation by redistributing gas and changing the spatial distribution of young stellar populations, thereby affecting wavelength-dependent surface brightness profiles \citep{2009_James, 2020_Fraser, 2020a_Nersesian, 2007_Wozniak}. 
The connection between stellar bars and star formation is complex, with observational studies reaching different conclusions. Bars can promote star formation by inducing gas inflows and concentrating gas in the central regions or along the bar \citep{2011Coelho, 2019Chown}. However, the presence of a bar does not necessarily lead to enhanced star formation, and suppression of star formation within the bar region has also been reported, particularly in massive or gas-poor galaxies \citep{2009Aguerri, 2015Willett, 2016James, 2018Khoperskov, 2020_Fraser, 2020Newnham}. 
Specifically, \citet{2020_Fraser} showed that bar-driven gas flows and angular momentum redistribution regulate the location of star formation along bars through shear, turbulence, and gas flows. Previous studies have shown that bars can redistribute young stellar populations, regulate the spatial distribution of star formation, and affect dust heating by young stars \citep{2007_Wozniak, 2009_James, 2020a_Nersesian}. These processes can produce a deficit of star formation within the bar region and enhanced star formation near the bar ends. These diverse observational results indicate that the effect of a bar on star formation is likely governed by the properties of its host galaxy and the evolutionary state of the bar \citep{2012Oh, 2012Masters, 2021Garcia, 2021Geron}.

Recently, a broken exponential radial surface brightness profile has been reported for the Milky Way, with the break radius located around 7.5~kpc or 1.3 $\rm R_{\rm e}$ (effective radii), significantly further inward than in typical Type II galaxies \citep{lian2024}. Within this break radius, the Milky Way exhibits a pronounced plateau in its surface brightness profile, a feature observed consistently across multiple stellar population ages. This behavior is highly unusual in the context of standard disk galaxy structures and challenges the long-standing view of the Milky Way as a simple exponential disk system.  
This revised structural picture has important implications for the inferred global properties of the Galaxy. The Milky Way has traditionally been considered relatively compact due to its short scale length \citep{vanderKruit2011,licquia2015,boardman2020_MWA}. However, measurements of the half-light radius based on the updated surface brightness profile suggest that the Milky Way has a size comparable to that of local galaxies with similar stellar masses \citep{lian2024}. Moreover, revised estimates of the total stellar mass, accounting for the plateau feature, yield a value of ${\rm 2.6\times10^{10}\ M_{\odot}}$ \citep{lian2025b}, approximately a factor of two lower than previous estimates based on a single exponential disk \citep{licquia2015,xiang2018}.  

The physical origin of the Milky Way’s plateau feature remains unclear. In particular, it is not yet established whether this structure arises primarily from stellar population variations, dynamical processes, or a combination of both. 
A plausible approach to addressing this problem is to study external galaxies that exhibit similar radial structures. Such systems span a broader range of physical conditions and allow direct measurements of radial profiles from an external perspective, thereby avoiding the line-of-sight complexities inherent to studies of the Milky Way.  

In this work, we systematically identify and analyze nearby galaxies hosting Milky Way–like plateau features. By combining multi-band surface brightness profiles and spectral energy distribution (SED) fitting, we aim to constrain the physical origin of the plateau and place the Milky Way within a broader extragalactic context. This paper is organized as follows. In Section~\ref{sec:SAMPLE}, we describe the sample selection and the procedures used to derive radial surface brightness profiles. In Section~\ref{sec:RESULT}, we present the multi-band surface brightness profiles, analyze radial color distributions, and perform SED fitting with \textsc{CIGALE} to derive stellar age and mass profiles. In Section~\ref{sec:DISCUSSION}, we discuss possible formation mechanisms of the plateau feature. Finally, a summary is provided in Section~\ref{sec:SUMMARY}. Throughout this work, we adopt the AB magnitude system.

\section{Data} \label{sec:SAMPLE}

To investigate the origin of the inner plateau in the Milky Way's radial structure, we identify and analyze external galaxies in the nearby universe that exhibit similar substructures. Our parent sample is drawn from a catalog of 15,424 nearby galaxies within 50~Mpc \citep{2024catalog}. We retrieve r-band imaging from the Dark Energy Spectroscopic Instrument (DESI) DR10 \citep{DESI_cite}. Due to incomplete sky coverage, this yields a working sample of 12,622 galaxies. Among these, 12,573 galaxies possess complete g, r, and z-band imaging (i.e., systems lacking one or more bands are excluded at this stage), which is required for subsequent spectral energy distribution (SED) analysis.  

To ensure data quality for reliable profile extraction, we perform flux validation on all 12,573 galaxies. For each image, we examine whether any pixel within a circular annulus centered on the catalog coordinates (RA, Dec) are saturated with flux exceeding 10 nanomaggies. The annulus is defined with an inner radius of $1/25$ and an outer radius of $2/5$ of the image size (in pixels). Galaxies satisfying this condition are excluded to remove contamination from bright neighboring sources, resulting in a cleaned sample of 12,008 galaxies. We then extract r-band radial surface brightness profiles to identify Milky Way–like systems exhibiting an inner plateau. The profiles are derived as following.  

\begin{figure}[ht]
    \centering
    \includegraphics[width=\textwidth]{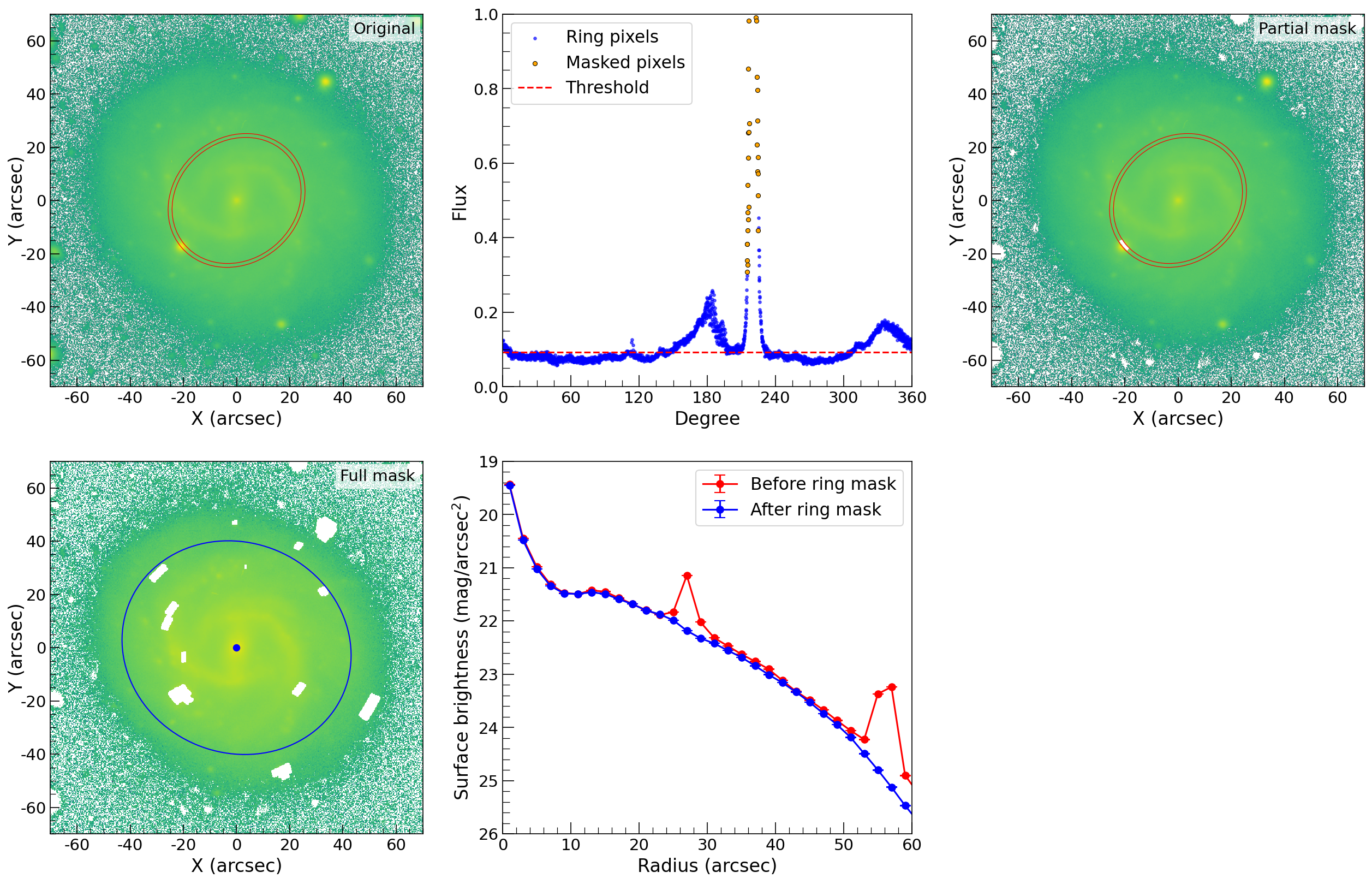}
    \caption{Illustration of the ring-mask procedure. 
    Top Left: Original image of a representative galaxy in our sample. The red elliptical annulus is defined by the measured axis ratio and position angle. 
    Top Middle: Flux distribution as a function of azimuthal angle within the annulus, measured relative to the galaxy center. The masking threshold is determined from the median and RMS of the flux distribution. 
    Top Right: Processed image after applying the ring mask at a specific radius. The foreground star in the lower-left region is effectively removed, demonstrating that the method suppresses localized contamination while preserving the intrinsic galaxy structure.
    Bottom Left: Final processed image obtained after applying the ring mask to the entire image. The blue elliptical annulus is defined by the axis ratio and position angle derived from the ring-masked image.
    Bottom Middle: Comparison of the surface brightness profiles before and after the ring-mask processing.
    }
    \label{fig:ring}
\end{figure}
 
We used the \texttt{photutils} package \citep{photutils} to detect sources, estimate the background, mask contaminating sources, and derive basic morphological parameters, including the galaxy center, axis ratio, and position angle. These parameters are subsequently used to construct a ring-mask for each galaxy to exclude foreground stars over the galaxy image. We generate a sequence of concentric elliptical annuli based on the derived morphological parameters. 
 Here, the position angle (PA) is determined from the overall morphology of each galaxy rather than the orientation of its bar. Since our aim is to identify and investigate plateaus in the radial surface brightness profiles of galaxies as a whole, rather than to characterize the internal radial structure of bars, we adopt the morphological PA throughout our analysis.
Within each annulus, we measure the pixel flux distribution and compute the median and root mean square (RMS). 
Masking is performed adaptively across different radial regimes. In the outer regions (i.e., beyond half of the diagonal length of the galaxy extent identified by \texttt{photutils}), where the flux dispersion becomes large and approaches the background level, we adopt a simplified masking scheme. The threshold is defined as the median flux within the annulus plus five times the RMS, and all pixels exceeding this threshold are directly masked. In the inner regions, the masking procedure is carried out iteratively along each elliptical annulus. We first identify high-flux pixels exceeding a threshold defined as the median plus five times the RMS. For each such pixel, its angular position relative to the galaxy center is determined, and an arc-shaped mask is constructed centered on that location. The angular extent of the arc is determined using a secondary reference level, defined as the median plus 0.1 times the RMS. We select pixels whose absolute deviation from this reference level is smaller than 0.1 times the median and treat them as approximate intersection points with this level. Along the annulus, the nearest such points on both sides of the high-flux pixel are identified and used as the angular boundaries of the mask. To avoid excessive masking caused by strong flux fluctuations, we impose an empirical upper limit on the angular extent of the arc, restricting it to within ±5 degrees around the high-flux pixel.
Figure~\ref{fig:ring} illustrates the ring mask for one annuli of a single galaxy. A full mask is obtained after looping through all annuli. This procedure ensures that localized contamination is minimized while preserving the intrinsic radial structure, which is critical for robust identification of shallow plateau features. The surface brightness profiles are then extracted from the masked images. 

We first apply this procedure to the r-band images. For all other bands, we adopt the r-band–derived center, axis ratio, and position angle to extract fluxes, ensuring consistent radial sampling across bands. Annular processing is also applied to the g, z bands from DESI, and u band from SDSS \citep[e.g.,][]{SDSS_cite,SDSS_cite_1}. For the near-infrared J-band images from Two Micron All Sky Survey (2MASS) \citep{2MASS_cite}, we do not perform masking due to its much lower image quality.

We identify Milky Way–like galaxies by searching for inner plateau features in their r-band radial surface brightness profiles. Guided by the Milky Way profile in \citet{lian2024}, we restrict our analysis to face-on galaxies (b/a $>$ 0.6) and focus on the radial range $0.3\,\rm R_e < R < 1.5\,R_e$ ($\rm R_e$ is the effective radius). Within this region, we identify segments with slopes flatter than 1 mag arcsec$^{-2}\ \rm R_e^{-1}$. A galaxy is classified as a candidate if such a segment spans more than $0.2\, \rm R_e$ and is followed by a region at larger radii where the slope steepens by at least a factor of two. This selection yields 306 candidate galaxies. From this preliminary sample, we visually select 122 galaxies with regular morphologies and no evidence of ongoing mergers.  
Since u-band data are required for reliable color gradients, we further restrict the sample to galaxies with available SDSS u-band imaging, resulting in a final sample of 65 galaxies.

To supplement the photometric data in some bands and assess the consistency between different datasets, we obtained additional multi-band observations with the Multi-channel Photometric Survey Telescope (Mephisto) \citep[e.g.,][]{yuan2020, chen2024sn}. The observations were conducted in the $u$, $v$, $g$, $r$ bands, covering wavelength ranges of 320--365, 365--405, 480--580, 580--680 nm, respectively. Mephisto is a 1.6 m wide-field multi-channel telescope and the first telescope of its kind in the world. It is located at the Lijiang Observatory of Yunnan Astronomical Observatory (IAU code: 044), at a longitude of $100^\circ01'48''$ E and a latitude of $26^\circ41'42''$ N, with an altitude of 3200 m. The telescope is operated by the South-Western Institute for Astronomy Research, Yunnan University. It is currently under commissioning and has already produced high-quality imaging data \citep[][e.g.,]{2024chena,2024yang}. The telescope is equipped with three CCD cameras, enabling simultaneous imaging in three channels: blue ($u$, $v$), yellow ($g$, $r$), and red ($i$, $z$). 
The reduction of the Mephisto data was carried out using a pipeline developed by the Mephisto team \citep{2024yang, 2025Du}.
Due to the observing schedule and time constraints, we obtained multi-band observations of seven galaxies from our sample with Mephisto. We examined the photometric data of these seven galaxies and found that the Mephisto $u$ and $r$ band measurements show good consistency with the corresponding SDSS $u$ band and DESI $r$ band measurements, respectively, with only minor differences between the datasets (see Figure~\ref{fig:diff_2}).

To investigate the possible role of bars in shaping the radial profiles, we divided the galaxies with available bar classifications into barred and non-barred systems using the Galaxy Zoo DESI Deep Learning Catalog \citep{galaxy_zoo}. The classification results are presented in Table~\ref{tab:params_a}. The cross-matching between catalogs was performed using a sky-coordinate matching radius of 5 arcsec. We used the sum of the \texttt{bar\_strong\_fraction} and \texttt{bar\_weak\_fraction} parameters, which represents the probability of a galaxy being classified as barred. 
 Following \cite{2021Geron}, a threshold of 0.5 was adopted to distinguish barred from non-barred galaxies, with galaxies having \texttt{bar\_strong\_fraction} + \texttt{bar\_weak\_fraction} $> 0.5$ classified as barred and those with values $\leq 0.5$ as non-barred. Galaxies without reliable \texttt{bar\_strong\_fraction} and \texttt{bar\_weak\_fraction} measurements are labeled as unclassified.
Based on this criterion, 25 galaxies are classified as barred and 17 as non-barred, while the remaining 23 galaxies lack corresponding classifications in the catalog. Thus, the barred fraction is approximately 60\% among the 42 galaxies with available classifications. 
The false-color images of the barred and non-barred galaxies in the MWL sample are shown in Figures A1 and A2, respectively.
For comparison, we constructed a control sample of 65 galaxies without plateau features in their $r$ band surface brightness profiles, matched to our sample in stellar mass and $g-r$ color distributions, and applied the same classification criterion to these galaxies.
The barred fraction in the control sample is approximately 32\% (based on 38 galaxies with available classifications). The higher barred fraction in the plateau sample suggests a possible statistical association between bars and plateau-like radial profiles, although the incomplete bar classifications and the presence of both barred and non-barred systems indicate that bars cannot by themselves account for all plateau features.

\begin{figure*} %[ht]
    \centering	
    \includegraphics[width=12cm]{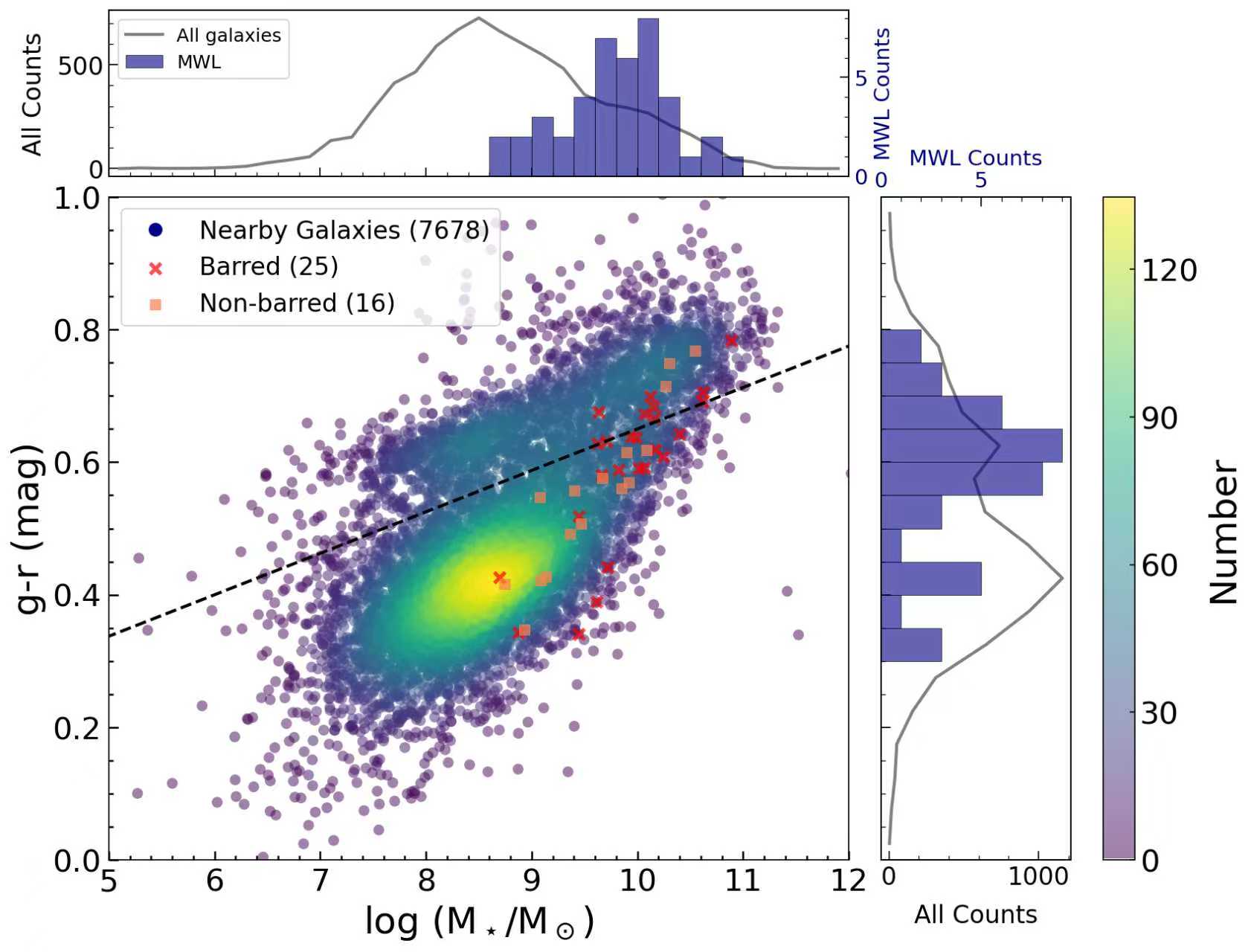}
    \caption{Distribution of stellar mass and color for the Milky Way–like sample of nearby galaxies (distance $< 50$~Mpc). 
    Red crosses and coral squares represent barred and non-barred galaxies, respectively.
    Stellar masses and g-r colors are taken from \citet{2024catalog}. 
    One non-barred galaxy has no corresponding information in this catalog and is therefore not shown; only the other 16 non-barred galaxies with available information are plotted.
    The density distribution of 7,678 nearby galaxies is included for reference. The black dashed line separates quiescent galaxies (upper region) from star-forming galaxies (lower region). Most galaxies in our sample lie near $\rm M_\ast \sim 10^{10}\,M_{\odot}$ and are predominantly star-forming.
    The top and right panels show the stellar mass and g-r color distributions of all galaxies and the MWL sample, respectively.
    }
    \label{fig:sample}
\end{figure*}

Figure~\ref{fig:sample} shows the distribution of our sample in the stellar mass--color plane. Stellar masses and g-r colors are adopted from the catalog of \citet{2024catalog}. One non-barred galaxies in our sample lack color measurements and are not included in the plot. For comparison, we also show the density distribution of 7,678 nearby galaxies ($< 50\ {\rm Mpc}$). The well-known bimodality between red quiescent and blue star-forming populations is clearly visible, and a dividing line is drawn based on visual inspection. 
Most galaxies in our sample have stellar masses above $10^{9}\, \rm M_{\odot}$ and are predominantly star-forming. 
Within this mass regime, Milky Way–like (MWL) systems constitute only $\sim3.6\%$ of the parent sample.

\section{RESULT} \label{sec:RESULT}
\subsection{Radial Surface Brightness Profiles} \label{subsec:Multiband variation}

\begin{figure}[ht]
    \centering
    \includegraphics[width=\textwidth]{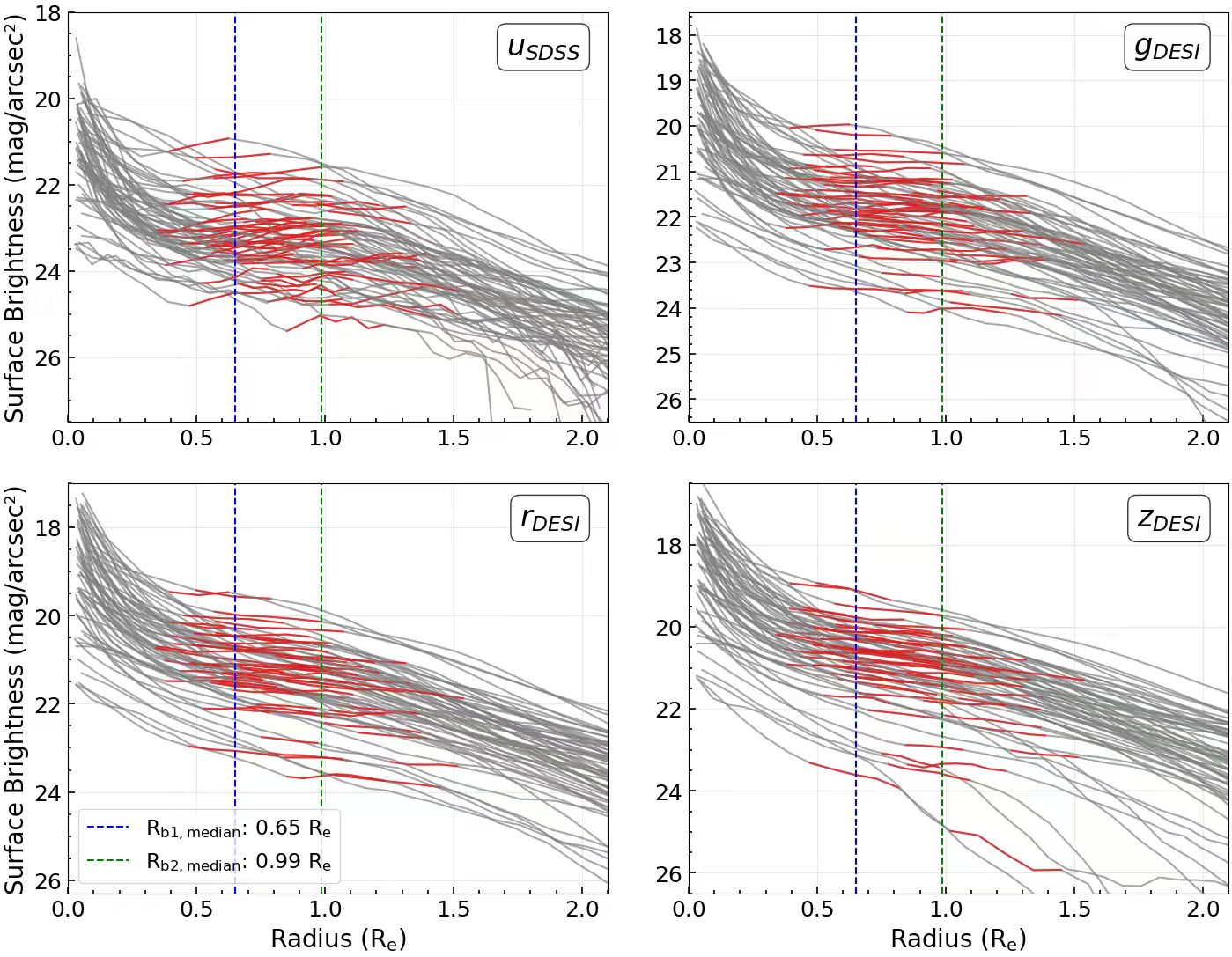}
    \caption{Radial surface brightness profiles in the u, g, r, and z bands. The red lines mark the plateau regions of individual galaxies, while the blue and green lines indicate the median inner and outer break radii in the r band, respectively. The deviation from an exponential profile becomes more pronounced toward shorter wavelengths.}
    \label{fig:profile}
\end{figure}

\begin{figure}[ht]
    \centering
    \includegraphics[width=\textwidth]{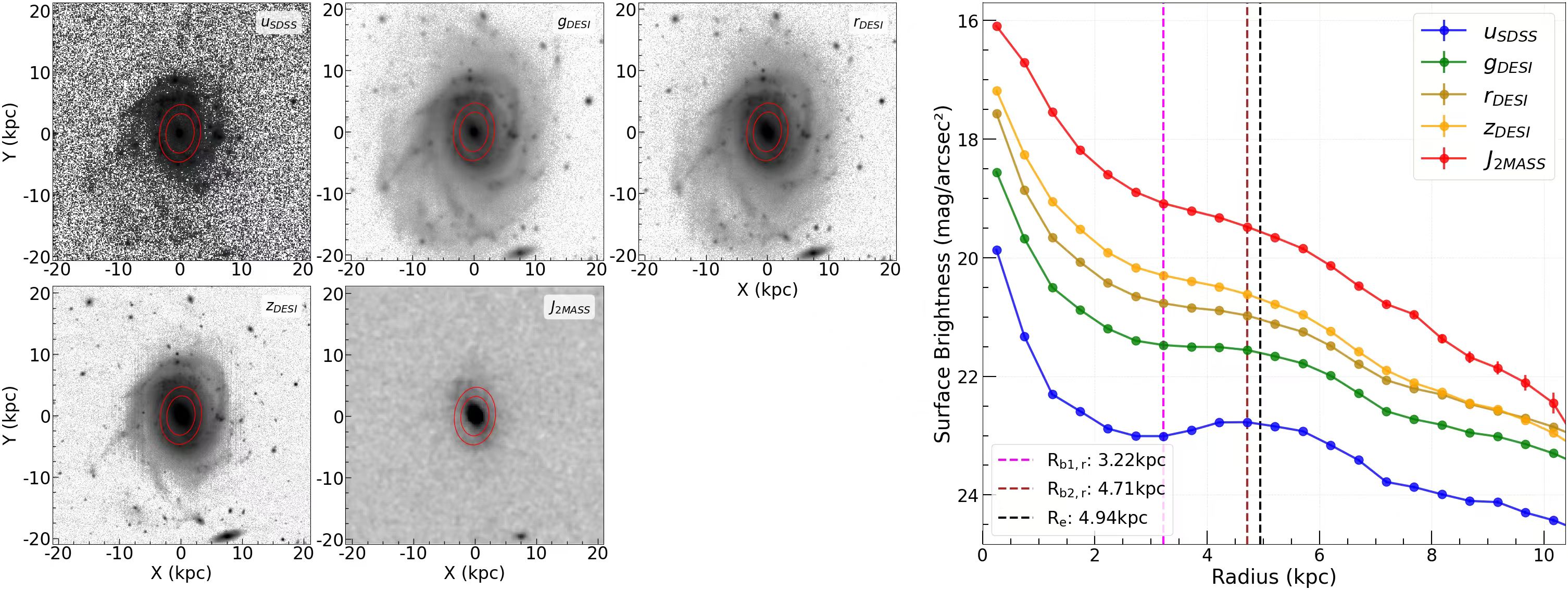}
    \caption{Left: Multi-band images ($u_{\mathrm{SDSS}}$, $g_{\mathrm{DESI}}$, $r_{\mathrm{DESI}}$, $z_{\mathrm{DESI}}$, $J_{\mathrm{2MASS}}$) of NGC 3811. The plateau region is indicated by two red elliptical annuli. 
    Right: Corresponding radial surface brightness profiles. The half-light radius ($\rm R_e$) is marked by a black dashed line, and the two r-band break radii are indicated by pink and brown dashed lines. A strong wavelength dependence is observed within the plateau region.}
    \label{fig:group_1}
\end{figure}

\begin{figure}[ht]
    \centering
    \includegraphics[width=\textwidth]{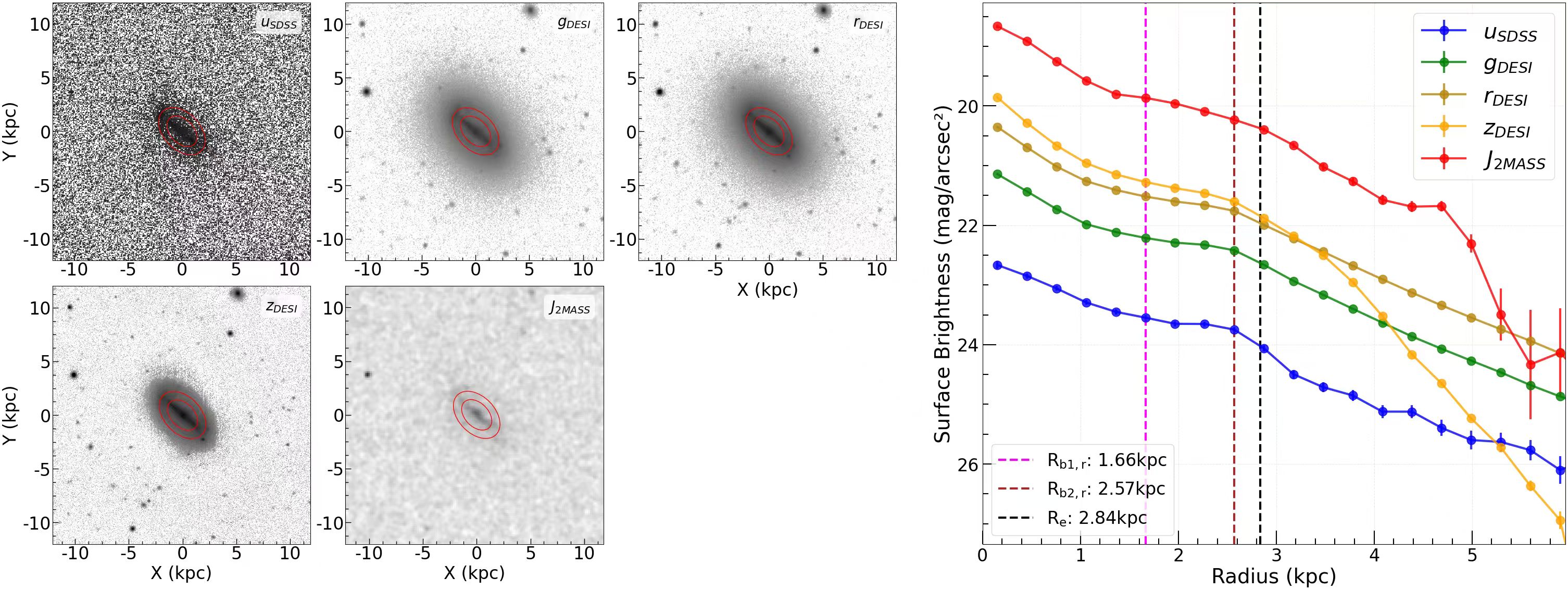}
    \caption{Same as Figure~\ref{fig:group_1}, but for UGC 08741. The plateau region shows minimal variation across different bands. A prominent bar structure is visible in the images.}
    \label{fig:group_2}
\end{figure}

Figure~\ref{fig:profile} presents the radial surface brightness profiles in multiple bands for the 65 galaxies in our sample. The plateau regions are highlighted in red, while the blue and green lines indicate the median inner and outer break radii in the r band, respectively. We observe a clear wavelength dependence in the plateau slopes: in the u band, the slope is typically negative, while it becomes progressively more negative in the redder bands. This behavior suggests that the plateau feature is not simply a structural component but is tightly coupled to stellar population properties.

To illustrate this trend, we examine two representative galaxies: NGC 3811, which shows a strong wavelength dependence, and UGC 08741, which exhibits little to no dependence. Their multi-band images and surface brightness profiles are shown in Figures~\ref{fig:group_1} and~\ref{fig:group_2}, respectively. The radial extent of the plateau is indicated by the pink (inner break) and brown (outer break) dashed lines in the profiles, and by the two red elliptical annuli overlaid on the images. For NGC 3811, the surface brightness profile shows a positive slope in the blue bands and a negative slope in the red bands within the plateau region, indicating strong wavelength dependence. In contrast, UGC 08741 shows nearly identical plateau slopes across all bands, suggesting minimal wavelength dependence. These contrasting behaviors highlight the complex nature of the plateau feature and suggest that different physical processes may be involved.

\begin{figure}[ht]
    \centering	
    \includegraphics[width=10cm]{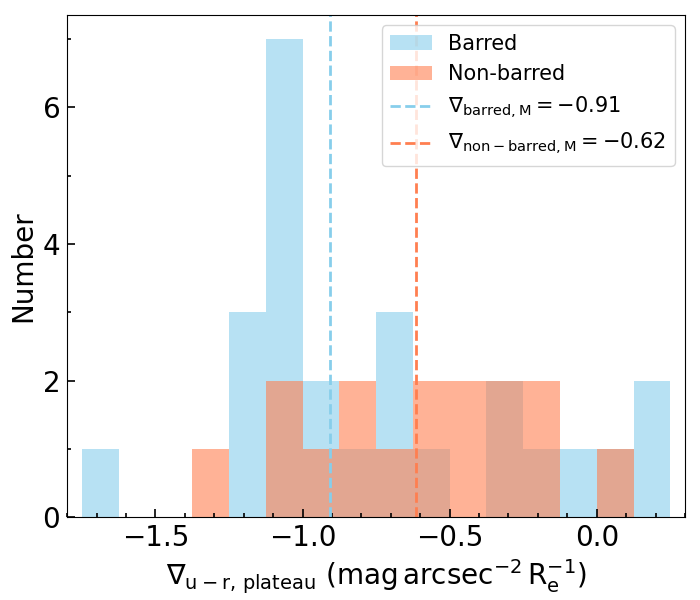}
    \caption{Distribution of color gradients within the plateau regions. Cyan and coral represent barred and non-barred galaxies, respectively, while the two dashed lines indicate their respective median values.}
    \label{fig:slope distribution}
\end{figure}

To quantify this wavelength-dependence, we analyze the radial color gradients within the plateau regions ($\nabla_{\mathrm {u - r, plateau}}$), the unit of this gradient is mag arcsec$^{-2}\ {\rm R_e^{-1}}$. 
In Figure~\ref{fig:slope distribution}, we show the distributions of the color gradients for the 42 galaxies with available bar classifications, divided into barred and non-barred systems. The barred galaxies tend to have steeper color gradients than the non-barred galaxies, as indicated by their more negative median value. However, both samples exhibit a broad range of color gradients. 
 As illustrated by the two representative galaxies in Figures \ref{fig:group_1} and \ref{fig:group_2}, NGC 3811 shows a strongly wavelength-dependent plateau (a steep color gradient), while UGC 08741 exhibits nearly identical plateau slopes across all bands (a shallow color gradient).
This suggests that the plateau profiles of at least some barred galaxies may be related to variations in their stellar population distributions. At the same time, the broad range of color gradients observed in both barred and non-barred galaxies indicates that the plateau features may have multiple origins rather than being produced by a single mechanism.

\subsection{SED Fitting} \label{subsec:Age and Mass analysis}

\begin{figure}[ht]
    \centering
    \includegraphics[width=\textwidth]{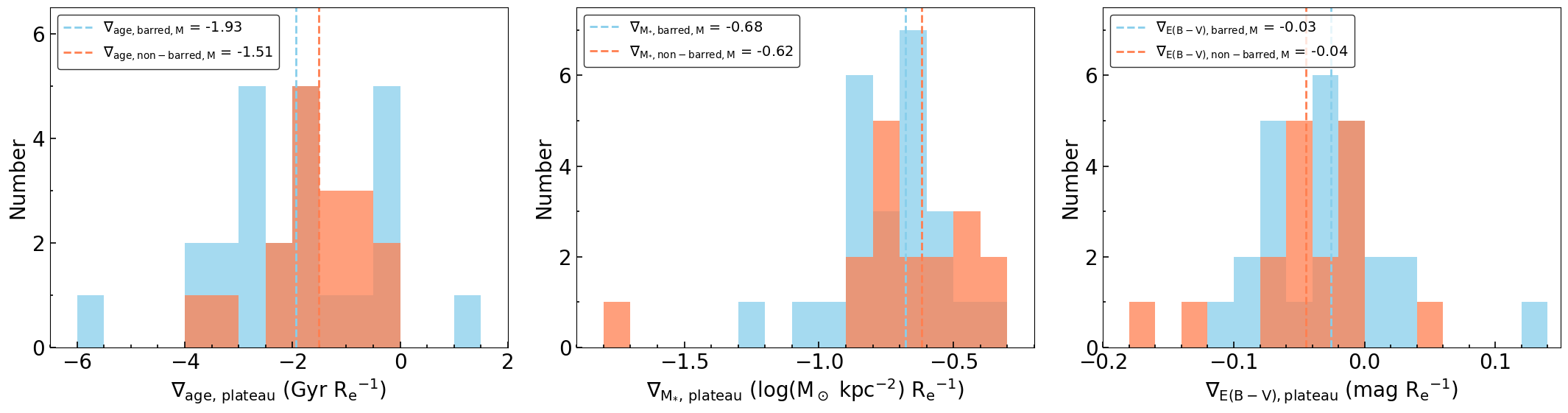}
    \caption{Distributions of radial gradients in stellar age, stellar mass and stellar color excess within the plateau regions. Cyan and coral represent barred and non-barred galaxies, respectively, while the corresponding dashed lines indicate their median values.}
    \label{fig:sed}
\end{figure}

To investigate the stellar population properties, we perform spectral energy distribution (SED) fitting using \textsc{CIGALE} \citep{2019cigale}. 
We perform the SED fitting independently for each radial bin to investigate the radial variations of the stellar population properties. It has been shown that broadband SED fitting can provide relatively robust constraints on stellar mass, while the recovery of stellar age and metallicity is more sensitive to the available wavelength coverage and model assumptions \citep{sed_1,sed_2,sed_3}. The adopted fitting parameters and their ranges are listed in Table~\ref{table:table}.  
The fitting incorporates multi-band photometric data from SDSS (u), DESI (g, r, z), and 2MASS (J, H, $\rm K_s$). For galaxies without full 2MASS coverage, only the optical bands are used. Among our 65 sample galaxies, 7 galaxies (NGC 2750, NGC 3811, NGC 5347, UGC 01862, UGC 06531, UGC 07138, UGC 08741) have been observed by Mephisto, providing additional u, v, g, and r band photometric data.

\begin{table}[ht!]
\centering
\small  
\setlength{\tabcolsep}{3pt} 
\caption{Model parameters used in the SED fitting}
\begin{tabularx}{\textwidth}{>{\centering\arraybackslash}X c c}
\hline\hline
Module  &  Parameter  & Values \\
\noalign{\smallskip}\hline
\multirow{3}{=}{Star formation history \newline (sfhdelayedbq\citep{2019cigale})}
& e-folding time of main population (Myr) 
& 1000,2000,3000,5000,8000 \\
& Age of main population (Myr) 
& 2000,5000,8000,10000 \\
& Age of burst/quench episode (Myr) 
& 5,10,20,50,100,300,500,1000 \\
\hline
\multirow{2}{=}{Simple stellar population \newline (bc03\citep{2003bc03})} 
& Initial mass function 
& 0 \\
& Metallicity 
& 0.02,0.05 \\
\hline
\multirow{2}{=}{Dust attenuation \newline (Calzetti et al. 2000\citep{dustatt})} 
& Color excess of nebular lines (mag) 
& 0.05,0.1,0.2, ... 0.9 \\
& Stellar continuum factor 
& 0.44 \\
\hline
\noalign{\smallskip}\hline
\end{tabularx}
\normalsize
\label{table:table}
\end{table}

Figure~\ref{fig:sed} shows the distributions of the stellar age, stellar mass, and 
stellar extinction gradients within the plateau regions for barred and non-barred galaxies. 
Both barred and non-barred galaxies exhibit broad distributions of gradients. 
Barred galaxies tend to have steeper age and flatter extinction gradients than non-barred galaxies. 
However, considering the uncertainties associated with broadband SED fitting, particularly for stellar age and dust attenuation, this difference should be regarded as a tentative trend rather than a statistically significant distinction.

\section{DISCUSSION} \label{sec:DISCUSSION}

In the following discussion, we consider several possible mechanisms that could contribute to the observed plateau features.

\subsection{Effects of bars}

\subsubsection{Bar-driven central homogeneous quenching} \label{subsec:Bar-driven central homogeneous quenching}

For barred galaxies, the stellar population gradients identified above may arise from the impact of bars on the central star formation history. In particular, bars can redistribute gas and regulate the spatial distribution of star formation, providing a possible pathway toward the suppression of star formation in the central regions. Previous studies have shown that bar-driven gas flows, angular momentum redistribution, shear, and turbulence can strongly affect the location and efficiency of star formation along the bar \citep{2020_Fraser, 2007_Wozniak}. For example, \citet{2009_James} found a deficit of $\mathrm{H}\alpha$ emission within bars, together with enhanced star formation toward the bar ends, suggesting that recent star formation can be suppressed over the bar region while remaining active in the surrounding disk. \citet{2007_Wozniak} similarly showed that newly formed stars can accumulate near the bar ends as they become trapped on elongated bar-supporting orbits. These results indicate that bars can substantially reshape the spatial distribution of young stellar populations.

One possible consequence of this process is central homogeneous quenching within the bar region. If bar-driven gas redistribution removes or depletes the gas available for star formation in the central region, the recent star formation rate can decrease over a relatively extended and spatially coherent region. The stellar population in this region would then gradually become older and redder, producing a strong radial gradient in stellar age and color between the quenched inner region and the actively star-forming outer disk. This interpretation is consistent with the reduced contribution of young stars found in barred regions by \citet{2020a_Nersesian}, who showed that the contribution of young stars to dust heating decreases to $\sim$53\% in the bar region and $\sim$38\% in the bulge.

Such a stellar population gradient could contribute to the wavelength-dependent plateau observed in some of our barred galaxies.
Because blue-band emission is more sensitive to recent star formation than red-band emission, the suppression of star formation in the central region causes a stronger reduction in blue-band surface brightness. The inner profile can therefore become relatively flat, or even positively sloped, in blue bands, while remaining negatively sloped in redder bands. 

This scenario also provides a possible explanation for the broad range of color gradients observed among barred galaxies. The strength and spatial extent of bar-driven gas redistribution and star formation suppression may vary from galaxy to galaxy, leading to different stellar population gradients and, consequently, different plateau profiles. 

\subsubsection{Bar-driven dynamical redistribution} \label{subsec:Bar-driven dynamical redistribution}

An alternative formation pathway for the plateau profiles in barred galaxies is bar-driven dynamical redistribution. Numerical simulations \citep{2021thor} have demonstrated that stellar bars can redistribute angular momentum and drive radial migration, transforming an initially exponential stellar mass surface density profile into a broken profile with a flattened inner region.

Unlike central quenching, which primarily alters the stellar population properties, dynamical redistribution directly changes the spatial distribution of stellar mass. Therefore, a plateau produced primarily through this mechanism is expected to show relatively weak wavelength dependence, since the redistribution affects the underlying stellar mass distribution rather than preferentially changing the contribution of young or old stellar populations. Such a process can produce a flattened surface brightness profile while leaving relatively weak color and stellar age gradients within the plateau region.

The efficiency of bar-driven redistribution may vary among galaxies depending on the properties and evolutionary state of the bar and the structure of the host galaxy. 
Consequently, dynamical redistribution may contribute to the observed plateau structure in some galaxies, but the present data do not allow us to determine whether this process dominates over star formation regulation or other mechanisms on a galaxy-by-galaxy basis.
The broad range of color gradients observed in our barred sample (Figure~\ref{fig:slope distribution}) is consistent with such a scenario, suggesting that different combinations of stellar population regulation and dynamical redistribution may contribute to the observed plateau features.

Finally, the Milky Way itself appears to exhibit characteristics of both formation channels. Its plateau shows a strong dependence on stellar population age, with younger populations displaying flatter slopes \citep{lian2024}. At the same time, the plateau persists in the stellar mass surface density profile, albeit with a steeper slope \citep{lian2025b}. 
Taken together, these observations suggest that the Milky Way’s plateau is unlikely to be explained by a single observable effect. Instead, both stellar population variations and changes in the stellar mass distribution may contribute to its observed radial structure.

\subsection{Dust extinction} \label{subsec:Dust extinction}

\begin{figure}[ht]
    \centering
    \includegraphics[width=1.0\textwidth]{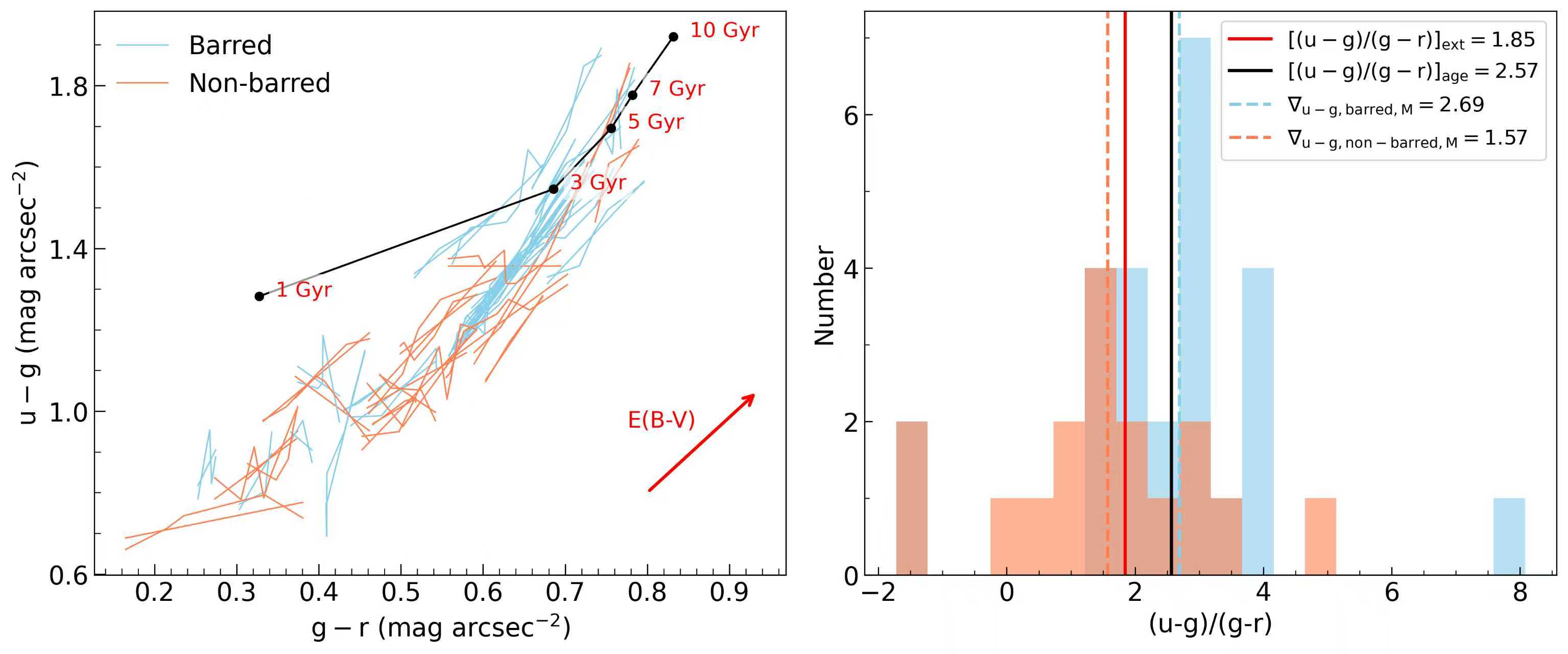}
    \caption{Left: Radial color evolution of the plateau regions in the $u-g$ versus $g-r$ color--color diagram. Each line represents one galaxy and traces its color variation with galactocentric radius across the plateau region. Cyan and coral lines indicate barred and non-barred galaxies, respectively. The red arrow shows the expected reddening vector predicted by the Milky Way dust extinction law, with labels indicating different $E(B-V)$ values. 
    The connected symbols show the intrinsic integrated colors of solar-metallicity PARSEC SSPs at ages of 1, 3, 5, 7, and 10~Gyr, weighted by the Kroupa initial mass function (IMF) and ordered by increasing age.
    Right: Distribution of the radial color-gradient ratios $\nabla_{\mathrm{u-g}}/\nabla_{\mathrm{g-r}}$ for barred and non-barred galaxies. Cyan and coral histograms represent barred and non-barred galaxies, respectively, with dashed lines indicating their median values. The solid red and black vertical lines show the expected color-gradient ratios from dust attenuation and stellar population ageing, respectively. The black line is derived from solar-metallicity SSP models with ages of 3, 5, 7, and 10~Gyr.}
    \label{fig:dust}
\end{figure}

To investigate whether dust is responsible for the plateau profile, we compare the color–color tracks of barred and non-barred galaxies across the plateau region. For each galaxy, the two colors measured at different radii within the plateau are connected by a line, showing how its colors vary with radius. The Milky Way dust extinction vector  and stellar population tracks of different ages are overlaid for comparison. 
We adopt the Galactic extinction coefficients of $A_u/E(B-V)=5.155$ for the SDSS $u$ band and $A_g/E(B-V)=3.214$ and $A_r/E(B-V)=2.165$ for the DESI $g$ and $r$ bands, respectively. Following the Galactic extinction law of Cardelli et al. \citep{cardelli1989}, these values yield $E(u-g)/E(g-r)=[A_u-A_g]/[A_g-A_r]\simeq1.85$, which indicates the expected color variation due to Galactic extinction.
To construct the SSP aging sequence, we adopt intrinsic PARSEC isochrones \citep{Bressan2012} with solar metallicity ($Z=0.0152$), zero extinction, and ages of 1, 3, 5, 7, and 10~Gyr. For each age, the stellar magnitudes along the isochrone are sampled as a function of initial stellar mass, and each mass interval is weighted by the expected number of stars predicted by the Kroupa initial mass function (IMF) \citep{Kroupa2001}. The stellar magnitudes are then converted to fluxes, and their IMF-weighted contributions are summed to obtain the integrated flux in each band. These fluxes are converted back to integrated magnitudes, from which the $u-g$ and $g-r$ colors are calculated, yielding one point for each SSP age in Figure~\ref{fig:dust}. 
We find that stellar population aging predicts a more rapid reddening in $u-g$ color than $g-r$ compared to that predicted by increasing dust attenuation. Moreover, the tracks of barred galaxies tend to have color variations align with the stellar population age sequence, whereas those of non-barred galaxies follow the Milky Way dust extinction vector.

We then fit the slopes of these color–color tracks, as shown in Figure~\ref{fig:dust}. The median slope of barred galaxies is close to the direction expected from stellar population aging, while that of non-barred galaxies is closer to the dust extinction vector. These results suggest that stellar population variations are likely an important contributor to the plateau profiles in barred galaxies, whereas dust attenuation may play a more important role in non-barred systems. This trend is qualitatively consistent with the slightly steeper dust gradients of non-barred galaxies than barred systems inferred from SED fitting as shown in Figure~\ref{fig:sed}. 
However, alignment with the dust or stellar population vectors does not uniquely distinguish between these effects because age, metallicity, and dust attenuation can produce partially degenerate color variations. We therefore regard dust attenuation as a plausible contributor rather than a unique explanation for the plateau profiles of non-barred galaxies. 

\subsection{Environment} \label{subsec:Environment}

\begin{figure}[ht]
    \centering
    \includegraphics[width=\textwidth]{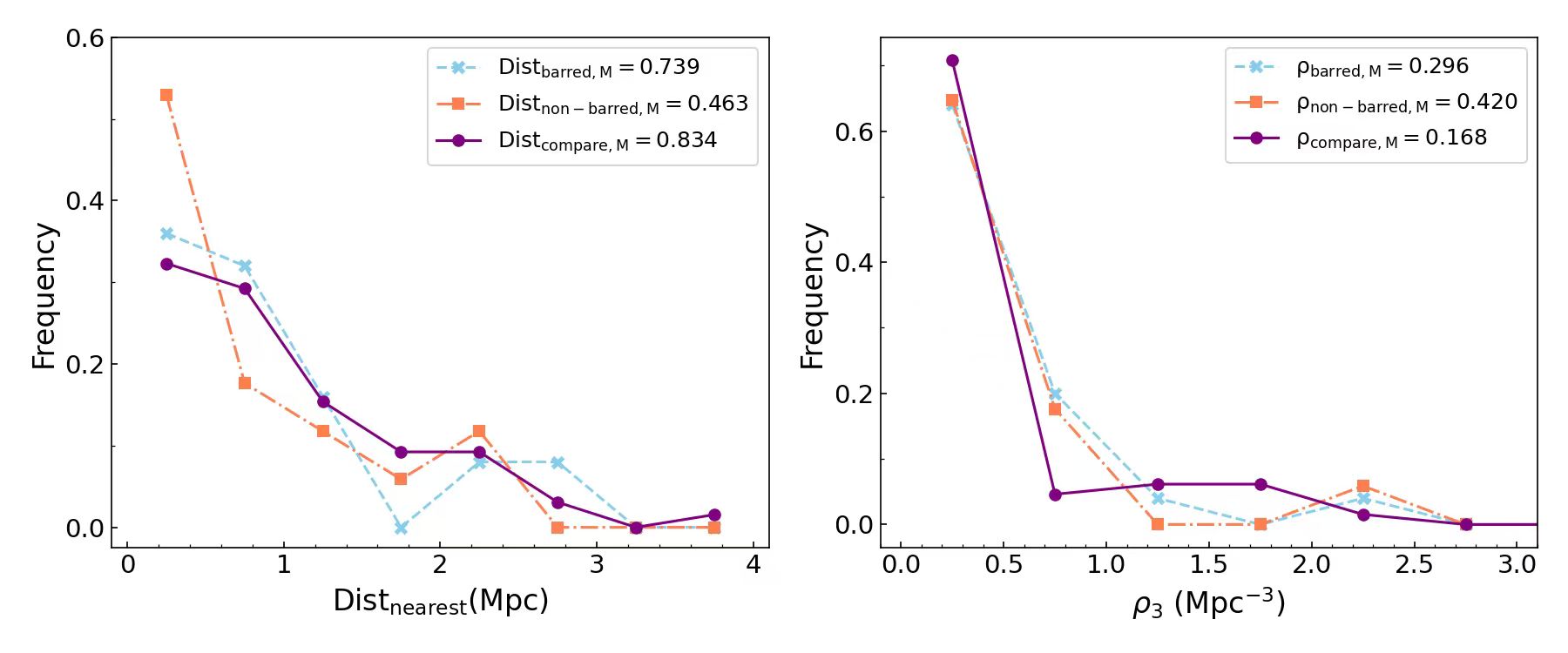}
    \caption{Left: Normalized distributions of the nearest-neighbor distance for barred (cyan; N = 25), non-barred (coral; N = 17), and comparison (purple; N = 65) galaxies. Right: Normalized distributions of the local galaxy number density, estimated from the distance to the third nearest neighbor. %Color coding is the same as in the left panel.
    }
    \label{fig:environment1}
\end{figure}

So far, we have focused primarily on the internal mechanisms that may contribute to the formation of plateau profiles. However, the environment in which a galaxy resides may also play an important role in shaping its radial structure. To investigate this possibility, we compare the environmental properties of the barred, non-barred, and control samples in Figure~\ref{fig:environment1}. We characterize the environment using two complementary indicators: the distance to the nearest neighboring galaxy and the local galaxy number density estimated from the third nearest neighbor. The nearest-neighbor distance provides a measure of the proximity of a galaxy to its closest neighboring system, while the third-nearest-neighbor density characterizes the local concentration of galaxies on a somewhat larger scale. Although these two indicators describe the environment in different ways, they show a consistent overall trend in our sample. The control galaxies generally reside in the lowest-density environments, followed by the barred plateau galaxies, while the non-barred plateau galaxies tend to inhabit the densest environments.

The consistency between the two environmental indicators strengthens the evidence that the difference in environmental properties between the plateau and control samples is not driven by a particular choice of environmental indicator. In particular, both indicators suggest that plateau galaxies preferentially reside in denser environments than the control galaxies, with the non-barred plateau galaxies showing the strongest tendency toward dense environments. 
These results suggest that environmental properties may be related to the presence of plateau profiles and that the environmental contribution may differ between barred and non-barred systems. However, the present analysis only reveals a correlation between environmental properties and the occurrence of plateau profiles and does not establish a causal connection between environmental density and plateau formation. Therefore, environmental effects should be regarded as a possible contributing factor rather than a uniquely identified mechanism for the formation of the observed plateaus.

\begin{sidewaystable}
\centering 
\small
\setlength{\tabcolsep}{3pt}
\caption{Parameters of our galaxy sample.}
\label{tab:params_a}
\begin{tabular}{lccccccccccc}
\hline
Objname & RA & Dec & Mass & Bar Type & $\mathrm{R_e}$ & $\mathrm{R_{b1}}$ & $\mathrm{R_{b2}}$ & $\nabla_{\rm mag,r}$ & $\nabla_{\rm u-r}$ & $\nabla_{\rm mass}$ & $\nabla_{\rm age}$ \\
& & & ($\log \mathrm{M_\odot}$) & & (kpc) & $\mathrm{(R_e)}$ & $\mathrm{(R_e)}$ & $(\mathrm{mag}\ \mathrm{arcsec}^{-2}\ \mathrm{R_\mathrm{e}^{-1}})$ & $(\mathrm{mag}\ \mathrm{arcsec}^{-2}\ \mathrm{R_\mathrm{e}^{-1}})$ & $(\log(\mathrm{M_\odot}) \, \mathrm{kpc}^{-2} \, \mathrm{R_\mathrm{e}^{-1}})$ & $(\mathrm{Gyr}/ \mathrm{R_e})$ \\
\hline
NGC 0157 & 8.69494 & -8.39650 & 10.26 & unclassified & 4.49 & 0.46 & 0.71 & $0.38 \pm 0.015$ & $-0.44 \pm 0.027$ & $-0.59 \pm 0.619$ & $-1.00 \pm 7.971$  \\
NGC 0575 & 22.69433 & 21.44042 & 9.87 & unclassified & 4.29 & 0.38 & 0.64 & $-0.01 \pm 0.016$ & $-0.69 \pm 0.125$ & $-0.43 \pm 0.636$ & $-1.47 \pm 8.582$  \\
PGC 005744 & 23.12298 & 4.60147 & 8.74 & non-barred & 1.82 & 1.13 & 1.39 & $0.48 \pm 0.018$ & $-1.02 \pm 0.086$ & $-0.84 \pm 0.610$ & $-0.08 \pm 3.756$  \\
NGC 0638 & 24.90771 & 7.23733 & 9.66 & barred & 2.16 & 0.58 & 1.02 & $0.66 \pm 0.007$ & $-0.73 \pm 0.068$ & $-0.58 \pm 0.676$ & $-1.28 \pm 7.361$  \\
IC 0211 & 32.78329 & 3.85247 & 9.71 & barred & 6.09 & 1.26 & 1.51 & $0.30 \pm 0.013$ & $0.11 \pm 0.254$ & $-0.32 \pm 0.949$ & $-0.42 \pm 7.064$  \\
UGC 01862 & 36.10334 & -2.16232 & 9.08 & non-barred & 1.74 & 0.56 & 0.99 & $0.60 \pm 0.009$ & $-0.57 \pm 0.079$ & $-0.51 \pm 0.514$ & $-2.01 \pm 5.971$  \\
NGC 1015 & 39.54822 & -1.31872 & 10.12 & barred & 3.19 & 0.82 & 1.11 & $0.39 \pm 0.009$ & $-0.08 \pm 0.111$ & $-0.43 \pm 0.535$ & $-0.27 \pm 8.414$  \\
NGC 1073 & 40.91902 & 1.37584 & 9.61 & unclassified & 4.26 & 0.53 & 0.78 & $0.27 \pm 0.008$ & $-1.33 \pm 0.039$ & $-0.94 \pm 0.451$ & $-3.28 \pm 5.157$  \\
PGC 012633 & 50.57284 & -7.09069 & 9.90 & non-barred & 2.64 & 0.73 & 1.12 & $0.48 \pm 0.009$ & $-1.34 \pm 0.044$ & $-0.83 \pm 0.401$ & $-3.21 \pm 5.327$  \\
NGC 2537 & 123.31129 & 45.99012 & 9.40 & non-barred & 1.32 & 0.49 & 0.94 & $0.09 \pm 0.006$ & $-0.69 \pm 0.012$ & $-0.46 \pm 0.291$ & $-1.55 \pm 3.170$  \\
NGC 2608 & 128.82158 & 28.47347 & 10.16 & barred & 4.21 & 0.83 & 1.04 & $0.05 \pm 0.036$ & $-0.86 \pm 0.066$ & $-0.63 \pm 0.722$ & $-1.56 \pm 9.641$  \\
NGC 2750 & 136.44968 & 25.43749 & 10.20 & unclassified & 6.65 & 0.92 & 1.23 & $0.09 \pm 0.016$ & $-0.80 \pm 0.064$ & $-0.59 \pm 0.457$ & $-1.79 \pm 4.461$  \\
NGC 2780 & 138.18491 & 34.92554 & 9.63 & barred & 2.22 & 0.35 & 0.91 & $0.36 \pm 0.005$ & $-0.31 \pm 0.018$ & $-0.53 \pm 0.188$ & $-0.55 \pm 2.598$  \\
NGC 3041 & 148.27977 & 16.67767 & 10.08 & unclassified & 5.43 & 0.44 & 0.75 & $0.62 \pm 0.009$ & $-0.43 \pm 0.031$ & $-0.65 \pm 0.358$ & $-0.90 \pm 4.246$  \\
PGC 029692 & 152.93641 & -2.73156 & 9.71 & barred & 2.37 & 0.54 & 0.84 & $0.64 \pm 0.015$ & $-1.20 \pm 0.192$ & $-0.78 \pm 0.474$ & $-1.88 \pm 6.744$  \\
NGC 3423 & 162.80973 & 5.84005 & 9.52 & unclassified & 3.56 & 1.05 & 1.37 & $0.55 \pm 0.007$ & $0.72 \pm 0.055$ & $-0.28 \pm 0.767$ & $-0.34 \pm 6.290$  \\
NGC 3433 & 163.01607 & 10.14828 & 10.35 & unclassified & 7.29 & 0.74 & 0.96 & $0.40 \pm 0.011$ & $-0.34 \pm 0.062$ & $-0.70 \pm 0.456$ & $-0.83 \pm 6.651$  \\
NGC 3485 & 165.00996 & 14.84153 & 8.33 & unclassified & 0.56 & 0.48 & 0.93 & $0.26 \pm 0.008$ & $-1.05 \pm 0.026$ & $-0.63 \pm 0.401$ & $-2.03 \pm 4.341$  \\
UGC 06309 & 169.44364 & 51.47661 & 10.07 & barred & 4.87 & 0.60 & 0.97 & $-0.16 \pm 0.013$ & $-1.01 \pm 0.061$ & $-0.63 \pm 0.543$ & $-2.98 \pm 6.578$  \\
NGC 3681 & 171.62420 & 16.86329 & 9.94 & barred & 1.88 & 0.73 & 0.94 & $0.68 \pm 0.006$ & $-0.64 \pm 0.012$ & $-1.28 \pm 0.200$ & $-3.33 \pm 3.564$  \\
NGC 3687 & 172.00245 & 29.51101 & 10.17 & barred & 3.37 & 0.53 & 0.76 & $0.71 \pm 0.017$ & $-1.11 \pm 0.033$ & $-0.83 \pm 0.483$ & $-1.91 \pm 5.754$  \\
UGC 06491 & 172.35027 & 34.87111 & 8.88 & barred & 3.80 & 1.02 & 1.45 & $0.67 \pm 0.011$ & $0.24 \pm 0.198$ & $0.69 \pm 0.515$ & $2.58 \pm 3.475$  \\
UGC 06531 & 173.20504 & 39.08493 & 9.00 & unclassified & 3.11 & 0.47 & 0.82 & $0.60 \pm 0.013$ & $-0.99 \pm 0.120$ & $-0.61 \pm 0.374$ & $-0.71 \pm 3.498$  \\
NGC 3729 & 173.45586 & 53.12585 & 9.98 & unclassified & 2.72 & 0.55 & 0.80 & $0.62 \pm 0.016$ & $-0.15 \pm 0.016$ & $-0.44 \pm 0.426$ & $-0.38 \pm 5.648$  \\
NGC 3811 & 175.31934 & 47.69083 & 10.40 & barred & 4.94 & 0.65 & 0.95 & $0.66 \pm 0.014$ & $-1.04 \pm 0.031$ & $-0.99 \pm 0.528$ & $-3.93 \pm 6.196$  \\
NGC 3876 & 176.36110 & 9.16078 & 9.61 & barred & 2.76 & 1.02 & 1.33 & $0.66 \pm 0.010$ & $0.17 \pm 0.021$ & $-0.85 \pm 0.187$ & $-0.44 \pm 1.005$  \\
PGC 037462 & 178.92755 & 32.18894 & 9.47 & non-barred & 1.58 & 0.58 & 1.07 & $0.44 \pm 0.001$ & $0.08 \pm 0.024$ & $-0.54 \pm 0.282$ & $-1.59 \pm 2.907$  \\
NGC 3982 & 179.11719 & 55.12526 & 9.92 & non-barred & 1.91 & 0.40 & 0.62 & $0.02 \pm 0.013$ & $-2.36 \pm 0.013$ & $-1.70 \pm 0.543$ & $-3.41 \pm 6.431$  \\
NGC 3985 & 179.17553 & 48.33395 & 9.30 & non-barred & 1.58 & 0.97 & 1.31 & $0.47 \pm 0.014$ & $-0.15 \pm 0.024$ & $-0.41 \pm 0.434$ & $-0.59 \pm 4.215$  \\
NGC 4025 & 179.79249 & 37.79349 & 9.57 & unclassified & 6.95 & 0.75 & 0.97 & $0.64 \pm 0.019$ & $-0.90 \pm 0.213$ & $-1.02 \pm 0.533$ & $0.07 \pm 2.144$  \\
NGC 4123 & 182.04634 & 2.87830 & 9.82 & barred & 4.21 & 0.65 & 0.87 & $0.31 \pm 0.015$ & $-1.03 \pm 0.069$ & $-0.76 \pm 0.488$ & $-3.08 \pm 7.003$  \\
\hline
\end{tabular}

\vspace{0.3cm}
\begin{minipage}{0.90\textwidth}
\footnotesize
\textbf{Note.} 
Columns (1)--(3) list the galaxy name and celestial coordinates.
Column (4) gives the stellar mass.
Column (5) gives the bar classification for each galaxy.
Column (6) presents the effective radius.
Columns (7) and (8) denote the inner and outer break radii normalized by $R_{\rm e}$.
Columns (9)--(12) give the radial gradients of the $r$-band surface brightness, $(u-r)$ color, stellar mass surface density, and stellar age, respectively.

\end{minipage}
\end{sidewaystable}

\begin{sidewaystable}
%\hspace*{-5cm}
\centering 
\addtocounter{table}{-1}
\small
\setlength{\tabcolsep}{3pt}
\caption{(Continued) Parameters of our galaxy sample.}
\label{tab:params_b}
\begin{tabular}{lccccccccccc}
\hline
Objname & RA & Dec & Mass & Bar Type & $\mathrm{R_e}$ & $\mathrm{R_{b1}}$ & $\mathrm{R_{b2}}$ & $\nabla_{\rm mag,r}$ & $\nabla_{\rm u-r}$ & $\nabla_{\rm mass}$ & $\nabla_{\rm age}$ \\
& & & ($\log \mathrm{M_\odot}$) & & (kpc) & $\mathrm{(R_e)}$ & $\mathrm{(R_e)}$ & $(\mathrm{mag}\ \mathrm{arcsec}^{-2}\ \mathrm{R_\mathrm{e}^{-1}})$ & $(\mathrm{mag}\ \mathrm{arcsec}^{-2}\ \mathrm{R_\mathrm{e}^{-1}})$ & $(\log(\mathrm{M_\odot}) \, \mathrm{kpc}^{-2} \, \mathrm{R_\mathrm{e}^{-1}})$ & $(\mathrm{Gyr}/ \mathrm{R_e})$ \\
\hline
UGC 07138 & 182.37005 & 26.22630 & 9.37 & non-barred & 4.68 & 0.84 & 1.06 & $0.13 \pm 0.030$ & $-0.36 \pm 0.257$ & $-0.54 \pm 0.841$ & $-1.43 \pm 7.104$  \\
NGC 4189 & 183.44706 & 13.42496 & 9.65 & unclassified & 2.59 & 0.34 & 1.02 & $0.72 \pm 0.006$ & $-0.78 \pm 0.026$ & $-0.76 \pm 0.151$ & $-3.27 \pm 2.134$  \\
NGC 4210 & 183.81601 & 65.98536 & 10.23 & unclassified & 4.56 & 0.42 & 0.69 & $0.64 \pm 0.014$ & $-1.34 \pm 0.052$ & $-0.82 \pm 0.503$ & $-3.89 \pm 7.140$  \\
NGC 4275 & 184.96913 & 27.62096 & 9.85 & non-barred & 1.73 & 0.57 & 0.99 & $0.52 \pm 0.007$ & $-1.07 \pm 0.013$ & $-0.75 \pm 0.307$ & $-1.52 \pm 3.338$  \\
NGC 4290 & 185.19801 & 58.09247 & 10.62 & barred & 5.69 & 0.80 & 1.14 & $0.77 \pm 0.010$ & $-1.12 \pm 0.051$ & $-0.73 \pm 0.557$ & $-2.55 \pm 7.241$  \\
NGC 4412 & 186.65028 & 3.96479 & 10.02 & barred & 3.18 & 0.69 & 1.01 & $0.16 \pm 0.009$ & $-0.33 \pm 0.019$ & $-0.67 \pm 0.472$ & $-1.80 \pm 4.960$  \\
NGC 4454 & 187.21145 & -1.93923 & 10.55 & non-barred & 4.66 & 0.68 & 0.98 & $0.83 \pm 0.004$ & $-0.48 \pm 0.033$ & $-0.73 \pm 0.335$ & $-1.59 \pm 5.538$  \\
NGC 4580 & 189.45158 & 5.36849 & 9.93 & unclassified & 2.32 & 0.40 & 0.61 & $0.43 \pm 0.016$ & $-1.59 \pm 0.047$ & $-0.93 \pm 0.690$ & $-5.34 \pm 10.011$  \\
PGC 158033 & 191.28781 & -6.31579 & 9.13 & non-barred & 2.20 & 0.69 & 1.06 & $0.67 \pm 0.005$ & $-0.95 \pm 0.109$ & $-0.77 \pm 0.638$ & $-1.75 \pm 5.268$  \\
NGC 4680 & 191.72813 & -11.63708 & 10.06 & unclassified & 2.73 & 0.45 & 0.84 & $0.41 \pm 0.001$ & $-0.47 \pm 0.015$ & $-0.54 \pm 0.441$ & $-0.77 \pm 4.903$  \\
NGC 5020 & 198.16622 & 12.59967 & 10.58 & unclassified & 7.73 & 0.56 & 0.81 & $0.57 \pm 0.012$ & $-0.69 \pm 0.042$ & $-0.64 \pm 0.443$ & $-1.35 \pm 5.357$  \\
ESO 576-032 & 199.96667 & -22.27884 & 9.69 & unclassified & 2.72 & 0.49 & 0.75 & $0.25 \pm 0.021$ & $-1.08 \pm 0.108$ & $-0.65 \pm 0.856$ & $-3.12 \pm 9.925$  \\
NGC 5147 & 201.58257 & 2.10055 & 9.09 & non-barred & 1.33 & 0.88 & 1.21 & $0.57 \pm 0.010$ & $-0.62 \pm 0.029$ & $-0.74 \pm 0.441$ & $-1.79 \pm 3.133$  \\
NGC 5240 & 203.97996 & 35.58825 & 9.85 & unclassified & 3.95 & 0.38 & 0.63 & $0.51 \pm 0.017$ & $-0.80 \pm 0.067$ & $-0.62 \pm 0.527$ & $-2.46 \pm 7.331$  \\
NGC 5248 & 204.38342 & 8.88518 & 10.26 & non-barred & 3.69 & 0.69 & 1.19 & $0.51 \pm 0.004$ & $-0.17 \pm 0.017$ & $-0.38 \pm 0.356$ & $-0.60 \pm 4.582$  \\
UGC 08741 & 207.23475 & 59.83588 & 9.71 & unclassified & 2.84 & 0.59 & 0.91 & $0.73 \pm 0.008$ & $-0.08 \pm 0.108$ & $-0.43 \pm 0.471$ & $-1.49 \pm 6.017$  \\
NGC 5325 & 207.72583 & 38.27478 & 9.45 & barred & 4.01 & 0.70 & 0.95 & $0.57 \pm 0.016$ & $-0.68 \pm 0.042$ & $-0.86 \pm 0.320$ & $-1.86 \pm 2.149$  \\
NGC 5347 & 208.32416 & 33.49082 & 10.16 & barred & 3.95 & 0.85 & 1.30 & $0.48 \pm 0.004$ & $-0.95 \pm 0.065$ & $-0.62 \pm 0.595$ & $-2.44 \pm 7.321$  \\
NGC 5339 & 208.50113 & -7.93061 & 10.24 & barred & 5.31 & 0.59 & 1.08 & $0.59 \pm 0.009$ & $-0.59 \pm 0.031$ & $-0.60 \pm 0.177$ & $-2.14 \pm 2.687$  \\
NGC 5378 & 209.21254 & 37.79722 & 10.44 & unclassified & 5.72 & 0.76 & 1.08 & $0.59 \pm 0.008$ & $-1.03 \pm 0.109$ & $-0.81 \pm 0.621$ & $-4.87 \pm 8.182$  \\
UGC 09123 & 213.76546 & 36.45730 & 8.70 & barred & 2.29 & 0.78 & 1.09 & $0.53 \pm 0.019$ & $-1.63 \pm 0.386$ & $-1.09 \pm 0.777$ & $0.77 \pm 4.234$  \\
NGC 5633 & 216.86820 & 46.14652 & 10.09 & non-barred & 2.42 & 0.50 & 0.79 & $0.65 \pm 0.001$ & $-0.82 \pm 0.010$ & $-0.85 \pm 0.369$ & $-2.12 \pm 4.856$  \\
IC 1067 & 223.27182 & 3.33176 & 9.62 & barred & 1.87 & 0.66 & 0.97 & $0.05 \pm 0.005$ & $-1.16 \pm 0.089$ & $-0.82 \pm 0.402$ & $-3.81 \pm 6.330$  \\
NGC 5850 & 226.78191 & 1.54455 & 10.89 & barred & 8.47 & 1.02 & 1.37 & $0.16 \pm 0.006$ & $-1.20 \pm 0.047$ & $-0.84 \pm 0.188$ & $-5.80 \pm 3.327$  \\
UGC 09741 & 227.13961 & 52.29616 & 9.45 & barred & 1.90 & 0.49 & 0.88 & $0.29 \pm 0.001$ & $-0.18 \pm 0.018$ & $-0.52 \pm 0.219$ & $-0.87 \pm 2.061$  \\
NGC 5905 & 228.84720 & 55.51719 & 10.62 & barred & 9.24 & 0.61 & 0.83 & $0.74 \pm 0.014$ & $-0.91 \pm 0.046$ & $-1.10 \pm 0.476$ & $-2.78 \pm 6.320$  \\
NGC 5957 & 233.84674 & 12.04757 & 9.99 & barred & 3.48 & 0.66 & 1.11 & $0.68 \pm 0.005$ & $-1.05 \pm 0.029$ & $-0.76 \pm 0.392$ & $-1.78 \pm 4.735$  \\
UGC 10290 & 243.63738 & 0.82169 & 8.93 & unclassified & 4.06 & 0.85 & 1.23 & $0.14 \pm 0.014$ & $-0.30 \pm 0.155$ & $-0.39 \pm 0.463$ & $-0.97 \pm 3.583$  \\
NGC 6267 & 254.53609 & 22.98510 & 10.07 & barred & 4.19 & 0.53 & 1.02 & $0.04 \pm 0.009$ & $-1.10 \pm 0.026$ & $-0.68 \pm 0.182$ & $-3.79 \pm 2.720$  \\
NGC 7137 & 327.05442 & 22.15953 & 9.67 & non-barred & 1.98 & 0.38 & 1.02 & $0.23 \pm 0.006$ & $-0.36 \pm 0.017$ & $-0.39 \pm 0.299$ & $-0.63 \pm 3.130$  \\
UGC 12208 & 342.52650 & -1.04925 & 9.67 & unclassified & 3.61 & 0.67 & 1.04 & $0.46 \pm 0.012$ & $-0.05 \pm 0.217$ & $-0.27 \pm 0.526$ & $-0.23 \pm 6.937$  \\
NGC 7428 & 344.33142 & -1.04904 & 10.30 & non-barred & 3.39 & 0.85 & 1.08 & $0.47 \pm 0.005$ & $-0.82 \pm 0.045$ & $-0.74 \pm 0.404$ & $-2.49 \pm 6.318$  \\
UGC 12648 & 352.97298 & -2.15609 & 8.93 & non-barred & 2.13 & 1.00 & 1.37 & $0.53 \pm 0.014$ & $-0.48 \pm 0.102$ & $-0.64 \pm 0.405$ & $-0.91 \pm 2.666$  \\
NGC 7743 & 356.08804 & 9.93403 & 10.20 & unclassified & 2.94 & 1.00 & 1.54 & $0.80 \pm 0.005$ & $0.05 \pm 0.065$ & $-0.32 \pm 0.345$ & $0.28 \pm 6.607$  \\
\hline
\end{tabular}
\end{sidewaystable}

\section{SUMMARY} \label{sec:SUMMARY}

In this work, we investigate the physical origin of the plateau feature in the Milky Way’s radial surface brightness profile by analyzing a sample of 65 nearby face-on galaxies that exhibit analogous structures. 
From a parent sample of 12,008 galaxies, we identify only 122 Milky Way–like (MWL) systems with plateau features. Most of these systems have stellar masses above $10^{9} \rm M_{\odot}$ and are predominantly star-forming, and within this stellar-mass regime, MWL systems constitute only $\sim3.6\%$ of the parent sample, implying that such plateau features are intrinsically rare in the local Universe.

Among the galaxies with available morphological classifications, 25 out of 42 ($\sim$60\%) are classified as barred, compared with $\sim$32\% in the control sample without a plateau. The higher barred fraction suggests that the presence of bars may be statistically associated with the occurrence of plateau profiles in our sample. We therefore divide the galaxies with available morphological classifications into barred (25) and non-barred (17) subsamples. %As shown in Figures~\ref{fig:dust}, 
The color variations within the plateaus of barred galaxies are more consistent with the direction expected from stellar population aging, whereas those of non-barred galaxies are more consistent with the dust attenuation. These differences suggest that stellar population variations and dust attenuation may contribute differently to the observed color gradients in barred and non-barred galaxies. A qualitatively consistent trend is revealed with SED fitting to the multi-band radial surface brightness profiles. However, the color--color tracks and SED fitting do not uniquely determine the physical origin of the plateau, and the possible contributions from stellar population evolution, dust attenuation, and dynamical processes may overlap.

For barred galaxies, the broad range of color gradients within the plateau regions indicates that the strength of the wavelength dependence varies substantially from galaxy to galaxy. 
This diversity suggests that bars may influence the inner radial structure through multiple pathways. Strong wavelength dependence may reflect substantial variations in stellar populations associated with changes in recent star formation, while weak wavelength dependence may be consistent with a greater contribution from changes in the underlying stellar mass distribution. 

Finally, we investigate whether the environment may also contribute to the formation or evolution of plateau profiles.  We find that the non-barred galaxies tend to reside in the densest environments, while barred galaxies occupy environments that are denser than those of the control sample. This environmental trend indicates that internal and environmental processes may both be relevant, but the present data do not establish a causal relation between environment and plateau formation.

The Milky Way itself exhibits characteristics of both classes: its plateau shows a dependence on stellar population age, while also persisting in the stellar mass surface density profile. 
This combination suggests that its plateau  
may reflect contributions from both stellar population variations and changes in the stellar mass distribution.

%% Please use the acknowledgment and contribution environments. This will 
%% be anonomyized when the "anonymous" style option is used. 
\begin{acknowledgments}

We sincerely thank the anonymous referee for the insightful and constructive comments and suggestions, which have helped us substantially improve the manuscript.

This work is supported by National Key R\&D Program of China (No. 2024YFA1611600), National Natural Science Foundation of China (No.12473021), Yunnan Province Science and Technology Department (No. 202105AE160021 and 202005AB160002), and Key Laboratory of Survey Science of Yunnan Province (No. 202449CE340002). 

This work makes use of data obtained with the Dark Energy Spectroscopic Instrument (DESI).DESI construction and operations is managed by the Lawrence Berkeley National Laboratory. This research is supported by the U.S. Department of Energy, Office of Science, Office of High-Energy Physics, under Contract No. DE–AC02–05CH11231, and by the National Energy Research Scientific Computing Center, a DOE Office of Science User Facility under the same contract. Additional support for DESI is provided by the U.S. National Science Foundation, Division of Astronomical Sciences under Contract No. AST-0950945 to the NSF’s National Optical-Infrared Astronomy Research Laboratory; the Science and Technology Facilities Council of the United Kingdom; the Gordon and Betty Moore Foundation; the Heising-Simons Foundation; the French Alternative Energies and Atomic Energy Commission (CEA); the National Council of Science and Technology of Mexico (CONACYT); the Ministry of Science and Innovation of Spain, and by the DESI Member Institutions. The DESI collaboration is honored to be permitted to conduct astronomical research on Iolkam Du’ag (Kitt Peak), a mountain with particular significance to the Tohono O’odham Nation.

Funding for SDSS-III has been provided by the Alfred P. Sloan Foundation, the Participating Institutions, the National Science Foundation, and the U.S. Department of Energy Office of Science. The SDSS-III web site is http://www.sdss3.org/.
SDSS-III is managed by the Astrophysical Research Consortium for the Participating Institutions of the SDSS-III Collaboration including the University of Arizona, the Brazilian Participation Group, Brookhaven National Laboratory, Carnegie Mellon University, University of Florida, the French Participation Group, the German Participation Group, Harvard University, the Instituto de Astrofisica de Canarias, the Michigan State/Notre Dame/JINA Participation Group, Johns Hopkins University, Lawrence Berkeley National Laboratory, Max Planck Institute for Astrophysics, Max Planck Institute for Extraterrestrial Physics, New Mexico State University, New York University, Ohio State University, Pennsylvania State University, University of Portsmouth, Princeton University, the Spanish Participation Group, University of Tokyo, University of Utah, Vanderbilt University, University of Virginia, University of Washington, and Yale University.

This publication makes use of data products from the Two Micron All Sky Survey, which is a joint project of the University of Massachusetts and the Infrared Processing and Analysis Center/California Institute of Technology, funded by the National Aeronautics and Space Administration and the National Science Foundation.

This research made use of Photutils, an Astropy package for
detection and photometry of astronomical sources \citep{photutils}.

Mephisto is developed at and operated by the 
South-Western Institute for Astronomy Research of 
Yunnan University (SWIFAR-YNU), funded by the 
``Yunnan University Development Plan for World-Class University'' and the ``Yunnan University Development Plan for World-Class Astronomy Discipline''.

\end{acknowledgments}

%% For this sample we use BibTeX plus aasjournalv7.bst to generate the

\bibliography{cite}

@ARTICLE{xiang2018,
       author = {{Xiang}, Maosheng and {Shi}, Jianrong and {Liu}, Xiaowei and {Yuan}, Haibo and {Chen}, Bingqiu and {Huang}, Yang and {Wang}, Chun and {Wu}, Yaqian and {Tian}, Zhijia and {Huo}, Zhiying and {Zhang}, Huawei and {Zhang}, Meng},
        title = "{Stellar Mass Distribution and Star Formation History of the Galactic Disk Revealed by Mono-age Stellar Populations from LAMOST}",
      journal = {\apjs},
         year = 2018,
        month = aug,
       volume = {237},
       number = {2},
          eid = {33},
        pages = {33},
          doi = {10.3847/1538-4365/aad237},
archivePrefix = {arXiv},
       eprint = {1807.04592},
 primaryClass = {astro-ph.GA},
       adsurl = {https://ui.adsabs.harvard.edu/abs/2018ApJS..237...33X}
}

@ARTICLE{licquia2015,
       author = {{Licquia}, Timothy C. and {Newman}, Jeffrey A.},
        title = "{Improved Estimates of the Milky Way's Stellar Mass and Star Formation Rate from Hierarchical Bayesian Meta-Analysis}",
      journal = {\apj},
         year = 2015,
        month = jun,
       volume = {806},
       number = {1},
          eid = {96},
        pages = {96},
          doi = {10.1088/0004-637X/806/1/96},
archivePrefix = {arXiv},
       eprint = {1407.1078},
 primaryClass = {astro-ph.GA},
       adsurl = {https://ui.adsabs.harvard.edu/abs/2015ApJ...806...96L}
}

@ARTICLE{boardman2020_MWA,
       author = {{Boardman}, N. and {Zasowski}, G. and {Seth}, A. and {Newman}, J. and {Andrews}, B. and {Bershady}, M. and {Bird}, J. and {Chiappini}, C. and {Fielder}, C. and {Fraser-McKelvie}, A. and {Jones}, A. and {Licquia}, T. and {Masters}, K.~L. and {Minchev}, I. and {Schiavon}, R.~P. and {Brownstein}, J.~R. and {Drory}, N. and {Lane}, R.~R.},
        title = "{Milky Way analogues in MaNGA: multiparameter homogeneity and comparison to the Milky Way}",
      journal = {\mnras},
         year = 2020,
        month = jan,
       volume = {491},
       number = {3},
        pages = {3672-3701},
          doi = {10.1093/mnras/stz3126},
archivePrefix = {arXiv},
       eprint = {1910.12896},
 primaryClass = {astro-ph.GA},
       adsurl = {https://ui.adsabs.harvard.edu/abs/2020MNRAS.491.3672B}
}

@ARTICLE{vanderKruit2011,
       author = {{van der Kruit}, P.~C. and {Freeman}, K.~C.},
        title = "{Galaxy Disks}",
      journal = {\araa},
         year = 2011,
        month = sep,
       volume = {49},
       number = {1},
        pages = {301-371},
          doi = {10.1146/annurev-astro-083109-153241},
archivePrefix = {arXiv},
       eprint = {1101.1771},
 primaryClass = {astro-ph.GA},
       adsurl = {https://ui.adsabs.harvard.edu/abs/2011ARA&A..49..301V}
}

@ARTICLE{2006Pohlen,
       author = {{Pohlen}, M. and {Trujillo}, I.},
        title = "{The structure of galactic disks. Studying late-type spiral galaxies using SDSS}",
      journal = {\aap},
         year = 2006,
        month = aug,
       volume = {454},
       number = {3},
        pages = {759-772},
          doi = {10.1051/0004-6361:20064883},
archivePrefix = {arXiv},
       eprint = {astro-ph/0603682},
 primaryClass = {astro-ph},
       adsurl = {https://ui.adsabs.harvard.edu/abs/2006A&A...454..759P}
}

@article{Zheng_2015,
doi = {10.1088/0004-637X/800/2/120},
url = {https://doi.org/10.1088/0004-637X/800/2/120},
year = {2015},
month = {feb},
publisher = {The American Astronomical Society},
volume = {800},
number = {2},
pages = {120},
author = {Zheng, Zheng and Thilker, David A. and Heckman, Timothy M. and Meurer, Gerhardt R. and Burgett, W. S. and Chambers, K. C. and Huber, M. E. and Kaiser, N. and Magnier, E. A. and Metcalfe, N. and Price, P. A. and Tonry, J. L. and Wainscoat, R. J. and Waters, C.},
title = {THE STRUCTURE AND STELLAR CONTENT OF THE OUTER DISKS OF GALAXIES: A NEW VIEW FROM THE Pan-STARRS1 MEDIUM DEEP SURVEY},
journal = {The Astrophysical Journal}
}

@ARTICLE{lian2024,
       author = {{Lian}, Jianhui and {Zasowski}, Gail and {Chen}, Bingqiu and {Imig}, Julie and {Wang}, Tao and {Boardman}, Nicholas and {Liu}, Xiaowei},
        title = "{The broken-exponential radial structure and larger size of the Milky Way galaxy}",
      journal = {Nature Astronomy},
         year = 2024,
        month = oct,
       volume = {8},
       number = {10},
        pages = {1302-1309},
          doi = {10.1038/s41550-024-02315-7},
archivePrefix = {arXiv},
       eprint = {2406.05604},
 primaryClass = {astro-ph.GA},
       adsurl = {https://ui.adsabs.harvard.edu/abs/2024NatAs...8.1302L}
}

@ARTICLE{2024catalog,
       author = {{Ohlson}, David and {Seth}, Anil C. and {Gallo}, Elena and {Baldassare}, Vivienne F. and {Greene}, Jenny E.},
        title = "{The 50 Mpc Galaxy Catalog (50 MGC): Consistent and Homogeneous Masses, Distances, Colors, and Morphologies}",
      journal = {\aj},
         year = 2024,
        month = jan,
       volume = {167},
       number = {1},
          eid = {31},
        pages = {31},
          doi = {10.3847/1538-3881/acf7bc},
archivePrefix = {arXiv},
       eprint = {2309.05701},
 primaryClass = {astro-ph.GA},
       adsurl = {https://ui.adsabs.harvard.edu/abs/2024AJ....167...31O}
}

@ARTICLE{2019cigale,
       author = {{Boquien}, M. and {Burgarella}, D. and {Roehlly}, Y. and {Buat}, V. and {Ciesla}, L. and {Corre}, D. and {Inoue}, A.~K. and {Salas}, H.},
        title = "{CIGALE: a python Code Investigating GALaxy Emission}",
      journal = {\aap},
         year = 2019,
        month = feb,
       volume = {622},
          eid = {A103},
        pages = {A103},
          doi = {10.1051/0004-6361/201834156},
archivePrefix = {arXiv},
       eprint = {1811.03094},
 primaryClass = {astro-ph.GA},
       adsurl = {https://ui.adsabs.harvard.edu/abs/2019A&A...622A.103B}
}

@ARTICLE{2003bc03,
       author = {{Bruzual}, G. and {Charlot}, S.},
        title = "{Stellar population synthesis at the resolution of 2003}",
      journal = {\mnras},
         year = 2003,
        month = oct,
       volume = {344},
       number = {4},
        pages = {1000-1028},
          doi = {10.1046/j.1365-8711.2003.06897.x},
archivePrefix = {arXiv},
       eprint = {astro-ph/0309134},
 primaryClass = {astro-ph},
       adsurl = {https://ui.adsabs.harvard.edu/abs/2003MNRAS.344.1000B}
}

@ARTICLE{2024yang,
       author = {{Yang}, Yuan-Pei and {Liu}, Xiangkun and {Pan}, Yu and {Er}, Xinzhong and {Liu}, Dezi and {Fang}, Yuan and {Du}, Guowang and {Cai}, Yongzhi and {Xu}, Xian and {Chen}, Xinlei and {Zou}, Xingzhu and {Guo}, Helong and {Liu}, Chenxu and {Cheng}, Yehao and {Kumar}, Brajesh and {Liu}, Xiaowei},
        title = "{Multiband Simultaneous Photometry of Type II SN 2023ixf with Mephisto and the Twin 50 cm Telescopes}",
      journal = {\apj},
         year = 2024,
        month = jul,
       volume = {969},
       number = {2},
          eid = {126},
        pages = {126},
          doi = {10.3847/1538-4357/ad4be3},
archivePrefix = {arXiv},
       eprint = {2405.08327},
 primaryClass = {astro-ph.HE},
       adsurl = {https://ui.adsabs.harvard.edu/abs/2024ApJ...969..126Y}
}

@ARTICLE{chen2024sn,
       author = {{Chen}, Xinlei and {Kumar}, Brajesh and {Er}, Xinzhong and {Guo}, Helong and {Yang}, Yuan-Pei and {Lin}, Weikang and {Fang}, Yuan and {Du}, Guowang and {Liu}, Chenxu and {Zhao}, Jiewei and {Zhang}, Tianyu and {Bao}, Yuxi and {Zou}, Xingzhu and {Pan}, Yu and {Wang}, Yu and {Zhu}, Xufeng and {Chatterjee}, Kaushik and {Liu}, Xiangkun and {Liu}, Dezi and {Lagioia}, Edoardo P. and {Rangwal}, Geeta and {Zhong}, Shiyan and {Zhang}, Jinghua and {Lian}, Jianhui and {Cai}, Yongzhi and {Zhang}, Yangwei and {Liu}, Xiaowei},
        title = "{Early-phase Simultaneous Multiband Observations of the Type II Supernova SN 2024ggi with Mephisto}",
      journal = {\apjl},
         year = 2024,
        month = aug,
       volume = {971},
       number = {1},
          eid = {L2},
        pages = {L2},
          doi = {10.3847/2041-8213/ad62f7},
archivePrefix = {arXiv},
       eprint = {2405.07964},
 primaryClass = {astro-ph.HE},
       adsurl = {https://ui.adsabs.harvard.edu/abs/2024ApJ...971L...2C}
}

@ARTICLE{lian2025b,
       author = {{Lian}, Jianhui and {Wang}, Tao and {Feng}, Qikang and {Huang}, Yang and {Guo}, Helong},
        title = "{The Milky Way Is a Less Massive Galaxy{\textemdash}New Estimates of the Milky Way's Local and Global Stellar Masses}",
      journal = {\apjl},
         year = 2025,
        month = sep,
       volume = {990},
       number = {2},
          eid = {L37},
        pages = {L37},
          doi = {10.3847/2041-8213/adfc73},
archivePrefix = {arXiv},
       eprint = {2508.13665},
 primaryClass = {astro-ph.GA},
       adsurl = {https://ui.adsabs.harvard.edu/abs/2025ApJ...990L..37L}
}

@ARTICLE{2021thor,
       author = {{Tepper-Garcia}, T. and {Bland-Hawthorn}, J. and {Vasiliev}, E. and {Athanassoula}, E. and {Gerhard}, O. and {Quillen}, A. and {McMillan}, P. and {Freeman}, K. and {Lewis}, G.~F. and {Teyssier}, R. and {Sharma}, S. and {Hayden}, M.~R. and {Buder}, S.},
        title = "{A barred Milky Way surrogate from an N-body simulation}",
      journal = {arXiv e-prints},
         year = 2021,
        month = nov,
          eid = {arXiv:2111.05466},
        pages = {arXiv:2111.05466},
          doi = {10.48550/arXiv.2111.05466},
archivePrefix = {arXiv},
       eprint = {2111.05466},
 primaryClass = {astro-ph.GA},
       adsurl = {https://ui.adsabs.harvard.edu/abs/2021arXiv211105466T}
}

@ARTICLE{2024chena,
       author = {{Chen}, Ping and {Gal-Yam}, Avishay and {Sollerman}, Jesper and {Schulze}, Steve and {Post}, Richard S. and {Liu}, Chang and {Ofek}, Eran O. and {Das}, Kaustav K. and {Fremling}, Christoffer and {Horesh}, Assaf and {Katz}, Boaz and {Kushnir}, Doron and {Kasliwal}, Mansi M. and {Kulkarni}, Shri R. and {Liu}, Dezi and {Liu}, Xiangkun and {Miller}, Adam A. and {Rose}, Kovi and {Waxman}, Eli and {Yang}, Sheng and {Yao}, Yuhan and {Zackay}, Barak and {Bellm}, Eric C. and {Dekany}, Richard and {Drake}, Andrew J. and {Fang}, Yuan and {Fynbo}, Johan P.~U. and {Groom}, Steven L. and {Helou}, George and {Irani}, Ido and {Jegou du Laz}, Theophile and {Liu}, Xiaowei and {Mazzali}, Paolo A. and {Neill}, James D. and {Qin}, Yu-Jing and {Riddle}, Reed L. and {Sharon}, Amir and {Strotjohann}, Nora L. and {Wold}, Avery and {Yan}, Lin},
        title = "{A 12.4-day periodicity in a close binary system after a supernova}",
      journal = {\nat},
         year = 2024,
        month = jan,
       volume = {625},
       number = {7994},
        pages = {253-258},
          doi = {10.1038/s41586-023-06787-x},
archivePrefix = {arXiv},
       eprint = {2310.07784},
 primaryClass = {astro-ph.HE},
       adsurl = {https://ui.adsabs.harvard.edu/abs/2024Natur.625..253C}
}

@ARTICLE{dustatt,
       author = {{Calzetti}, Daniela and {Armus}, Lee and {Bohlin}, Ralph C. and {Kinney}, Anne L. and {Koornneef}, Jan and {Storchi-Bergmann}, Thaisa},
        title = "{The Dust Content and Opacity of Actively Star-forming Galaxies}",
      journal = {\apj},
         year = 2000,
        month = apr,
       volume = {533},
       number = {2},
        pages = {682-695},
          doi = {10.1086/308692},
archivePrefix = {arXiv},
       eprint = {astro-ph/9911459},
 primaryClass = {astro-ph},
       adsurl = {https://ui.adsabs.harvard.edu/abs/2000ApJ...533..682C}
}

@ARTICLE{DESI_cite,
       author = {{Dey}, Arjun and {Schlegel}, David J. and {Lang}, Dustin and {Blum}, Robert and {Burleigh}, Kaylan and {Fan}, Xiaohui and {Findlay}, Joseph R. and {Finkbeiner}, Doug and {Herrera}, David and {Juneau}, St{\'e}phanie and {Landriau}, Martin and {Levi}, Michael and {McGreer}, Ian and {Meisner}, Aaron and {Myers}, Adam D. and {Moustakas}, John and {Nugent}, Peter and {Patej}, Anna and {Schlafly}, Edward F. and {Walker}, Alistair R. and {Valdes}, Francisco and {Weaver}, Benjamin A. and {Y{\`e}che}, Christophe and {Zou}, Hu and {Zhou}, Xu and {Abareshi}, Behzad and {Abbott}, T.~M.~C. and {Abolfathi}, Bela and {Aguilera}, C. and {Alam}, Shadab and {Allen}, Lori and {Alvarez}, A. and {Annis}, James and {Ansarinejad}, Behzad and {Aubert}, Marie and {Beechert}, Jacqueline and {Bell}, Eric F. and {BenZvi}, Segev Y. and {Beutler}, Florian and {Bielby}, Richard M. and {Bolton}, Adam S. and {Brice{\~n}o}, C{\'e}sar and {Buckley-Geer}, Elizabeth J. and {Butler}, Karen and {Calamida}, Annalisa and {Carlberg}, Raymond G. and {Carter}, Paul and {Casas}, Ricard and {Castander}, Francisco J. and {Choi}, Yumi and {Comparat}, Johan and {Cukanovaite}, Elena and {Delubac}, Timoth{\'e}e and {DeVries}, Kaitlin and {Dey}, Sharmila and {Dhungana}, Govinda and {Dickinson}, Mark and {Ding}, Zhejie and {Donaldson}, John B. and {Duan}, Yutong and {Duckworth}, Christopher J. and {Eftekharzadeh}, Sarah and {Eisenstein}, Daniel J. and {Etourneau}, Thomas and {Fagrelius}, Parker A. and {Farihi}, Jay and {Fitzpatrick}, Mike and {Font-Ribera}, Andreu and {Fulmer}, Leah and {G{\"a}nsicke}, Boris T. and {Gaztanaga}, Enrique and {George}, Koshy and {Gerdes}, David W. and {Gontcho}, Satya Gontcho A. and {Gorgoni}, Claudio and {Green}, Gregory and {Guy}, Julien and {Harmer}, Diane and {Hernandez}, M. and {Honscheid}, Klaus and {Huang}, Lijuan Wendy and {James}, David J. and {Jannuzi}, Buell T. and {Jiang}, Linhua and {Joyce}, Richard and {Karcher}, Armin and {Karkar}, Sonia and {Kehoe}, Robert and {Kneib}, Jean-Paul and {Kueter-Young}, Andrea and {Lan}, Ting-Wen and {Lauer}, Tod R. and {Le Guillou}, Laurent and {Le Van Suu}, Auguste and {Lee}, Jae Hyeon and {Lesser}, Michael and {Perreault Levasseur}, Laurence and {Li}, Ting S. and {Mann}, Justin L. and {Marshall}, Robert and {Mart{\'\i}nez-V{\'a}zquez}, C.~E. and {Martini}, Paul and {du Mas des Bourboux}, H{\'e}lion and {McManus}, Sean and {Meier}, Tobias Gabriel and {M{\'e}nard}, Brice and {Metcalfe}, Nigel and {Mu{\~n}oz-Guti{\'e}rrez}, Andrea and {Najita}, Joan and {Napier}, Kevin and {Narayan}, Gautham and {Newman}, Jeffrey A. and {Nie}, Jundan and {Nord}, Brian and {Norman}, Dara J. and {Olsen}, Knut A.~G. and {Paat}, Anthony and {Palanque-Delabrouille}, Nathalie and {Peng}, Xiyan and {Poppett}, Claire L. and {Poremba}, Megan R. and {Prakash}, Abhishek and {Rabinowitz}, David and {Raichoor}, Anand and {Rezaie}, Mehdi and {Robertson}, A.~N. and {Roe}, Natalie A. and {Ross}, Ashley J. and {Ross}, Nicholas P. and {Rudnick}, Gregory and {Safonova}, Sasha and {Saha}, Abhijit and {S{\'a}nchez}, F. Javier and {Savary}, Elodie and {Schweiker}, Heidi and {Scott}, Adam and {Seo}, Hee-Jong and {Shan}, Huanyuan and {Silva}, David R. and {Slepian}, Zachary and {Soto}, Christian and {Sprayberry}, David and {Staten}, Ryan and {Stillman}, Coley M. and {Stupak}, Robert J. and {Summers}, David L. and {Sien Tie}, Suk and {Tirado}, H. and {Vargas-Maga{\~n}a}, Mariana and {Vivas}, A. Katherina and {Wechsler}, Risa H. and {Williams}, Doug and {Yang}, Jinyi and {Yang}, Qian and {Yapici}, Tolga and {Zaritsky}, Dennis and {Zenteno}, A. and {Zhang}, Kai and {Zhang}, Tianmeng and {Zhou}, Rongpu and {Zhou}, Zhimin},
        title = "{Overview of the DESI Legacy Imaging Surveys}",
      journal = {\aj},
         year = 2019,
        month = may,
       volume = {157},
       number = {5},
          eid = {168},
        pages = {168},
          doi = {10.3847/1538-3881/ab089d},
archivePrefix = {arXiv},
       eprint = {1804.08657},
 primaryClass = {astro-ph.IM},
       adsurl = {https://ui.adsabs.harvard.edu/abs/2019AJ....157..168D}
}

@ARTICLE{SDSS_cite,
       author = {{Ahn}, Christopher P. and {Alexandroff}, Rachael and {Allende Prieto}, Carlos and {Anderson}, Scott F. and {Anderton}, Timothy and {Andrews}, Brett H. and {Aubourg}, {\'E}ric and {Bailey}, Stephen and {Balbinot}, Eduardo and {Barnes}, Rory and {Bautista}, Julian and {Beers}, Timothy C. and {Beifiori}, Alessandra and {Berlind}, Andreas A. and {Bhardwaj}, Vaishali and {Bizyaev}, Dmitry and {Blake}, Cullen H. and {Blanton}, Michael R. and {Blomqvist}, Michael and {Bochanski}, John J. and {Bolton}, Adam S. and {Borde}, Arnaud and {Bovy}, Jo and {Brandt}, W.~N. and {Brinkmann}, J. and {Brown}, Peter J. and {Brownstein}, Joel R. and {Bundy}, Kevin and {Busca}, N.~G. and {Carithers}, William and {Carnero}, Aurelio R. and {Carr}, Michael A. and {Casetti-Dinescu}, Dana I. and {Chen}, Yanmei and {Chiappini}, Cristina and {Comparat}, Johan and {Connolly}, Natalia and {Crepp}, Justin R. and {Cristiani}, Stefano and {Croft}, Rupert A.~C. and {Cuesta}, Antonio J. and {da Costa}, Luiz N. and {Davenport}, James R.~A. and {Dawson}, Kyle S. and {de Putter}, Roland and {De Lee}, Nathan and {Delubac}, Timoth{\'e}e and {Dhital}, Saurav and {Ealet}, Anne and {Ebelke}, Garrett L. and {Edmondson}, Edward M. and {Eisenstein}, Daniel J. and {Escoffier}, S. and {Esposito}, Massimiliano and {Evans}, Michael L. and {Fan}, Xiaohui and {Femen{\'\i}a Castell{\'a}}, Bruno and {Fern{\'a}ndez Alvar}, Emma and {Ferreira}, Leticia D. and {Filiz Ak}, N. and {Finley}, Hayley and {Fleming}, Scott W. and {Font-Ribera}, Andreu and {Frinchaboy}, Peter M. and {Garc{\'\i}a-Hern{\'a}ndez}, D.~A. and {Garc{\'\i}a P{\'e}rez}, A.~E. and {Ge}, Jian and {G{\'e}nova-Santos}, R. and {Gillespie}, Bruce A. and {Girardi}, L{\'e}o and {Gonz{\'a}lez Hern{\'a}ndez}, Jonay I. and {Grebel}, Eva K. and {Gunn}, James E. and {Guo}, Hong and {Haggard}, Daryl and {Hamilton}, Jean-Christophe and {Harris}, David W. and {Hawley}, Suzanne L. and {Hearty}, Frederick R. and {Ho}, Shirley and {Hogg}, David W. and {Holtzman}, Jon A. and {Honscheid}, Klaus and {Huehnerhoff}, J. and {Ivans}, Inese I. and {Ivezi{\'c}}, {\v{Z}}eljko and {Jacobson}, Heather R. and {Jiang}, Linhua and {Johansson}, Jonas and {Johnson}, Jennifer A. and {Kauffmann}, Guinevere and {Kirkby}, David and {Kirkpatrick}, Jessica A. and {Klaene}, Mark A. and {Knapp}, Gillian R. and {Kneib}, Jean-Paul and {Le Goff}, Jean-Marc and {Leauthaud}, Alexie and {Lee}, Khee-Gan and {Lee}, Young Sun and {Long}, Daniel C. and {Loomis}, Craig P. and {Lucatello}, Sara and {Lundgren}, Britt and {Lupton}, Robert H. and {Ma}, Bo and {Ma}, Zhibo and {MacDonald}, Nicholas and {Mack}, Claude E. and {Mahadevan}, Suvrath and {Maia}, Marcio A.~G. and {Majewski}, Steven R. and {Makler}, Martin and {Malanushenko}, Elena and {Malanushenko}, Viktor and {Manchado}, A. and {Mandelbaum}, Rachel and {Manera}, Marc and {Maraston}, Claudia and {Margala}, Daniel and {Martell}, Sarah L. and {McBride}, Cameron K. and {McGreer}, Ian D. and {McMahon}, Richard G. and {M{\'e}nard}, Brice and {Meszaros}, Sz. and {Miralda-Escud{\'e}}, Jordi and {Montero-Dorta}, Antonio D. and {Montesano}, Francesco and {Morrison}, Heather L. and {Muna}, Demitri and {Munn}, Jeffrey A. and {Murayama}, Hitoshi and {Myers}, Adam D. and {Neto}, A.~F. and {Nguyen}, Duy Cuong and {Nichol}, Robert C. and {Nidever}, David L. and {Noterdaeme}, Pasquier and {Nuza}, Sebasti{\'a}n E. and {Ogando}, Ricardo L.~C. and {Olmstead}, Matthew D. and {Oravetz}, Daniel J. and {Owen}, Russell and {Padmanabhan}, Nikhil and {Palanque-Delabrouille}, Nathalie and {Pan}, Kaike and {Parejko}, John K. and {Parihar}, Prachi and {P{\^a}ris}, Isabelle and {Pattarakijwanich}, Petchara and {Pepper}, Joshua and {Percival}, Will J. and {P{\'e}rez-Fournon}, Ismael and {P{\'e}rez-R{\`a}fols}, Ignasi and {Petitjean}, Patrick and {Pforr}, Janine and {Pieri}, Matthew M. and {Pinsonneault}, Marc H. and {Porto de Mello}, G.~F. and {Prada}, Francisco and {Price-Whelan}, Adrian M. and {Raddick}, M. Jordan and {Rebolo}, Rafael and {Rich}, James and {Richards}, Gordon T. and {Robin}, Annie C. and {Rocha-Pinto}, Helio J. and {Rockosi}, Constance M. and {Roe}, Natalie A. and {Ross}, Ashley J. and {Ross}, Nicholas P. and {Rossi}, Graziano and {Rubi{\~n}o-Martin}, J.~A. and {Samushia}, Lado and {Sanchez Almeida}, J. and {S{\'a}nchez}, Ariel G. and {Santiago}, Bas{\'\i}lio and {Sayres}, Conor and {Schlegel}, David J. and {Schlesinger}, Katharine J. and {Schmidt}, Sarah J. and {Schneider}, Donald P. and {Schultheis}, Mathias and {Schwope}, Axel D. and {Sc{\'o}ccola}, C.~G. and {Seljak}, Uros and {Sheldon}, Erin and {Shen}, Yue and {Shu}, Yiping and {Simmerer}, Jennifer and {Simmons}, Audrey E. and {Skibba}, Ramin A. and {Skrutskie}, M.~F. and {Slosar}, A. and {Sobreira}, Flavia and {Sobeck}, Jennifer S. and {Stassun}, Keivan G. and {Steele}, Oliver and {Steinmetz}, Matthias},
        title = "{The Ninth Data Release of the Sloan Digital Sky Survey: First Spectroscopic Data from the SDSS-III Baryon Oscillation Spectroscopic Survey}",
      journal = {\apjs},
         year = 2012,
        month = dec,
       volume = {203},
       number = {2},
          eid = {21},
        pages = {21},
          doi = {10.1088/0067-0049/203/2/21},
archivePrefix = {arXiv},
       eprint = {1207.7137},
 primaryClass = {astro-ph.IM},
       adsurl = {https://ui.adsabs.harvard.edu/abs/2012ApJS..203...21A}
}

@ARTICLE{SDSS_cite_1,
       author = {{York}, Donald G. and {Adelman}, J. and {Anderson}, Jr., John E. and {Anderson}, Scott F. and {Annis}, James and {Bahcall}, Neta A. and {Bakken}, J.~A. and {Barkhouser}, Robert and {Bastian}, Steven and {Berman}, Eileen and {Boroski}, William N. and {Bracker}, Steve and {Briegel}, Charlie and {Briggs}, John W. and {Brinkmann}, J. and {Brunner}, Robert and {Burles}, Scott and {Carey}, Larry and {Carr}, Michael A. and {Castander}, Francisco J. and {Chen}, Bing and {Colestock}, Patrick L. and {Connolly}, A.~J. and {Crocker}, J.~H. and {Csabai}, Istv{\'a}n and {Czarapata}, Paul C. and {Davis}, John Eric and {Doi}, Mamoru and {Dombeck}, Tom and {Eisenstein}, Daniel and {Ellman}, Nancy and {Elms}, Brian R. and {Evans}, Michael L. and {Fan}, Xiaohui and {Federwitz}, Glenn R. and {Fiscelli}, Larry and {Friedman}, Scott and {Frieman}, Joshua A. and {Fukugita}, Masataka and {Gillespie}, Bruce and {Gunn}, James E. and {Gurbani}, Vijay K. and {de Haas}, Ernst and {Haldeman}, Merle and {Harris}, Frederick H. and {Hayes}, J. and {Heckman}, Timothy M. and {Hennessy}, G.~S. and {Hindsley}, Robert B. and {Holm}, Scott and {Holmgren}, Donald J. and {Huang}, Chi-hao and {Hull}, Charles and {Husby}, Don and {Ichikawa}, Shin-Ichi and {Ichikawa}, Takashi and {Ivezi{\'c}}, {\v{Z}}eljko and {Kent}, Stephen and {Kim}, Rita S.~J. and {Kinney}, E. and {Klaene}, Mark and {Kleinman}, A.~N. and {Kleinman}, S. and {Knapp}, G.~R. and {Korienek}, John and {Kron}, Richard G. and {Kunszt}, Peter Z. and {Lamb}, D.~Q. and {Lee}, B. and {Leger}, R. French and {Limmongkol}, Siriluk and {Lindenmeyer}, Carl and {Long}, Daniel C. and {Loomis}, Craig and {Loveday}, Jon and {Lucinio}, Rich and {Lupton}, Robert H. and {MacKinnon}, Bryan and {Mannery}, Edward J. and {Mantsch}, P.~M. and {Margon}, Bruce and {McGehee}, Peregrine and {McKay}, Timothy A. and {Meiksin}, Avery and {Merelli}, Aronne and {Monet}, David G. and {Munn}, Jeffrey A. and {Narayanan}, Vijay K. and {Nash}, Thomas and {Neilsen}, Eric and {Neswold}, Rich and {Newberg}, Heidi Jo and {Nichol}, R.~C. and {Nicinski}, Tom and {Nonino}, Mario and {Okada}, Norio and {Okamura}, Sadanori and {Ostriker}, Jeremiah P. and {Owen}, Russell and {Pauls}, A. George and {Peoples}, John and {Peterson}, R.~L. and {Petravick}, Donald and {Pier}, Jeffrey R. and {Pope}, Adrian and {Pordes}, Ruth and {Prosapio}, Angela and {Rechenmacher}, Ron and {Quinn}, Thomas R. and {Richards}, Gordon T. and {Richmond}, Michael W. and {Rivetta}, Claudio H. and {Rockosi}, Constance M. and {Ruthmansdorfer}, Kurt and {Sandford}, Dale and {Schlegel}, David J. and {Schneider}, Donald P. and {Sekiguchi}, Maki and {Sergey}, Gary and {Shimasaku}, Kazuhiro and {Siegmund}, Walter A. and {Smee}, Stephen and {Smith}, J. Allyn and {Snedden}, S. and {Stone}, R. and {Stoughton}, Chris and {Strauss}, Michael A. and {Stubbs}, Christopher and {SubbaRao}, Mark and {Szalay}, Alexander S. and {Szapudi}, Istvan and {Szokoly}, Gyula P. and {Thakar}, Anirudda R. and {Tremonti}, Christy and {Tucker}, Douglas L. and {Uomoto}, Alan and {Vanden Berk}, Dan and {Vogeley}, Michael S. and {Waddell}, Patrick and {Wang}, Shu-i. and {Watanabe}, Masaru and {Weinberg}, David H. and {Yanny}, Brian and {Yasuda}, Naoki and {SDSS Collaboration}},
        title = "{The Sloan Digital Sky Survey: Technical Summary}",
      journal = {\aj},
         year = 2000,
        month = sep,
       volume = {120},
       number = {3},
        pages = {1579-1587},
          doi = {10.1086/301513},
archivePrefix = {arXiv},
       eprint = {astro-ph/0006396},
 primaryClass = {astro-ph},
       adsurl = {https://ui.adsabs.harvard.edu/abs/2000AJ....120.1579Y}
}

@ARTICLE{2MASS_cite,
       author = {{Skrutskie}, M.~F. and {Cutri}, R.~M. and {Stiening}, R. and {Weinberg}, M.~D. and {Schneider}, S. and {Carpenter}, J.~M. and {Beichman}, C. and {Capps}, R. and {Chester}, T. and {Elias}, J. and {Huchra}, J. and {Liebert}, J. and {Lonsdale}, C. and {Monet}, D.~G. and {Price}, S. and {Seitzer}, P. and {Jarrett}, T. and {Kirkpatrick}, J.~D. and {Gizis}, J.~E. and {Howard}, E. and {Evans}, T. and {Fowler}, J. and {Fullmer}, L. and {Hurt}, R. and {Light}, R. and {Kopan}, E.~L. and {Marsh}, K.~A. and {McCallon}, H.~L. and {Tam}, R. and {Van Dyk}, S. and {Wheelock}, S.},
        title = "{The Two Micron All Sky Survey (2MASS)}",
      journal = {\aj},
         year = 2006,
        month = feb,
       volume = {131},
       number = {2},
        pages = {1163-1183},
          doi = {10.1086/498708},
       adsurl = {https://ui.adsabs.harvard.edu/abs/2006AJ....131.1163S}
}

@ARTICLE{environment,
       author = {{Mondelin}, M. and {Bournaud}, F. and {Cuillandre}, J.-C. and {Codis}, S. and {Stone}, C. and {Bolzonella}, M. and {Sorce}, J.~G. and {Kluge}, M. and {Hatch}, N.~A. and {Marleau}, F.~R. and {Schirmer}, M. and {Bouy}, H. and {Buitrago}, F. and {Tortora}, C. and {Quilley}, L. and {George}, K. and {Baes}, M. and {Saifollahi}, T. and {Sanchez-Alarcon}, P.~M. and {Knapen}, J.~H. and {Aghanim}, N. and {Amara}, A. and {Andreon}, S. and {Baccigalupi}, C. and {Balestra}, A. and {Bardelli}, S. and {Battaglia}, P. and {Biviano}, A. and {Branchini}, E. and {Brescia}, M. and {Brinchmann}, J. and {Capobianco}, V. and {Carbone}, C. and {Castellano}, M. and {Castignani}, G. and {Cavuoti}, S. and {Cimatti}, A. and {Congedo}, G. and {Conselice}, C.~J. and {Conversi}, L. and {Copin}, Y. and {Courbin}, F. and {Courtois}, H.~M. and {Cropper}, M. and {De Lucia}, G. and {Dupac}, X. and {Fabricius}, M. and {Farina}, M. and {Faustini}, F. and {Ferriol}, S. and {Fotopoulou}, S. and {Gillis}, B. and {Giocoli}, C. and {Grupp}, F. and {Haugan}, S.~V.~H. and {Holmes}, W. and {Hormuth}, F. and {Hornstrup}, A. and {Jahnke}, K. and {Jhabvala}, M. and {Keih{\"a}nen}, E. and {Kermiche}, S. and {Kilbinger}, M. and {Kubik}, B. and {K{\"u}mmel}, M. and {Kurki-Suonio}, H. and {Le Brun}, A.~M.~C. and {Ligori}, S. and {Lilje}, P.~B. and {Lindholm}, V. and {Lloro}, I. and {Maino}, D. and {Maiorano}, E. and {Mansutti}, O. and {Marcin}, S. and {Marggraf}, O. and {Martinelli}, M. and {Medinaceli}, E. and {Mellier}, Y. and {Merlin}, E. and {Meylan}, G. and {Moscardini}, L. and {Niemi}, S.-M. and {Padilla}, C. and {Pasian}, F. and {Pedersen}, K. and {Percival}, W.~J. and {Pettorino}, V. and {Pires}, S. and {Poncet}, M. and {Popa}, L.~A. and {Pozzetti}, L. and {Renzi}, A. and {Riccio}, G. and {Romelli}, E. and {Saglia}, R. and {Schneider}, P. and {Secroun}, A. and {Serrano}, S. and {Sirignano}, C. and {Steinwagner}, J. and {Tereno}, I. and {Toledo-Moreo}, R. and {Torradeflot}, F. and {Tutusaus}, I. and {Valenziano}, L. and {Vassallo}, T. and {Wang}, Y. and {Weller}, J. and {Zerbi}, F.~M. and {Zucca}, E. and {Burigana}, C. and {Scottez}, V.},
        title = "{Euclid: Early Release Observations ─ The surface brightness and colour profiles of the far outskirts of galaxies in the Perseus cluster}",
      journal = {\aap},
         year = 2025,
        month = jul,
       volume = {699},
          eid = {A214},
        pages = {A214},
          doi = {10.1051/0004-6361/202554838},
archivePrefix = {arXiv},
       eprint = {2506.02745},
 primaryClass = {astro-ph.GA},
       adsurl = {https://ui.adsabs.harvard.edu/abs/2025A&A...699A.214M}
}

@ARTICLE{Xu_Yu_2024,
       author = {{Xu}, Dewang and {Yu}, Si-Yue},
        title = "{JWST reveals a high fraction of disk breaks at 1 {\ensuremath{\leq}} z {\ensuremath{\leq}} 3}",
      journal = {\aap},
         year = 2024,
        month = feb,
       volume = {682},
          eid = {L17},
        pages = {L17},
          doi = {10.1051/0004-6361/202449252},
archivePrefix = {arXiv},
       eprint = {2402.04233},
 primaryClass = {astro-ph.GA},
       adsurl = {https://ui.adsabs.harvard.edu/abs/2024A&A...682L..17X}
}

@ARTICLE{up_bending,
       author = {{Watkins}, Aaron E. and {Laine}, Jarkko and {Comer{\'o}n}, S{\'e}bastien and {Janz}, Joachim and {Salo}, Heikki},
        title = "{Varied origins of up-bending breaks in galaxy disks}",
      journal = {\aap},
         year = 2019,
        month = may,
       volume = {625},
          eid = {A36},
        pages = {A36},
          doi = {10.1051/0004-6361/201935130},
archivePrefix = {arXiv},
       eprint = {1903.09384},
 primaryClass = {astro-ph.GA},
       adsurl = {https://ui.adsabs.harvard.edu/abs/2019A&A...625A..36W}
}

@ARTICLE{down-bending_inside-out,
       author = {{Chen}, Liufei and {Du}, Min and {Lu}, Shuai and {Li}, Jing and {Ho}, Luis C.},
        title = "{Down-bending Breaks in Galactic Disks Are an Intrinsic Byproduct of Inside-out Growth}",
      journal = {arXiv e-prints},
         year = 2026,
        month = jan,
          eid = {arXiv:2602.00626},
        pages = {arXiv:2602.00626},
          doi = {10.48550/arXiv.2602.00626},
archivePrefix = {arXiv},
       eprint = {2602.00626},
 primaryClass = {astro-ph.GA},
       adsurl = {https://ui.adsabs.harvard.edu/abs/2026arXiv260200626C}
}

@ARTICLE{1970_Freeman,
       author = {{Freeman}, K.~C.},
        title = "{On the Disks of Spiral and S0 Galaxies}",
      journal = {\apj},
         year = 1970,
        month = jun,
       volume = {160},
        pages = {811},
          doi = {10.1086/150474},
       adsurl = {https://ui.adsabs.harvard.edu/abs/1970ApJ...160..811F}
}

@ARTICLE{1980Fall,
       author = {{Fall}, S.~M. and {Efstathiou}, G.},
        title = "{Formation and rotation of disc galaxies with haloes.}",
      journal = {\mnras},
         year = 1980,
        month = oct,
       volume = {193},
        pages = {189-206},
          doi = {10.1093/mnras/193.2.189},
       adsurl = {https://ui.adsabs.harvard.edu/abs/1980MNRAS.193..189F}
}

@ARTICLE{1989Kennicutt,
       author = {{Kennicutt}, Jr., Robert C.},
        title = "{The Star Formation Law in Galactic Disks}",
      journal = {\apj},
         year = 1989,
        month = sep,
       volume = {344},
        pages = {685},
          doi = {10.1086/167834},
       adsurl = {https://ui.adsabs.harvard.edu/abs/1989ApJ...344..685K}
}

@ARTICLE{2004Schaye,
       author = {{Schaye}, Joop},
        title = "{Star Formation Thresholds and Galaxy Edges: Why and Where}",
      journal = {\apj},
         year = 2004,
        month = jul,
       volume = {609},
       number = {2},
        pages = {667-682},
          doi = {10.1086/421232},
archivePrefix = {arXiv},
       eprint = {astro-ph/0205125},
 primaryClass = {astro-ph},
       adsurl = {https://ui.adsabs.harvard.edu/abs/2004ApJ...609..667S}
}

@ARTICLE{2006Elmegreen,
       author = {{Elmegreen}, Bruce G. and {Hunter}, Deidre A.},
        title = "{Radial Profiles of Star Formation in the Far Outer Regions of Galaxy Disks}",
      journal = {\apj},
         year = 2006,
        month = jan,
       volume = {636},
       number = {2},
        pages = {712-720},
          doi = {10.1086/498082},
archivePrefix = {arXiv},
       eprint = {astro-ph/0509190},
 primaryClass = {astro-ph},
       adsurl = {https://ui.adsabs.harvard.edu/abs/2006ApJ...636..712E}
}

@ARTICLE{2012Martin,
       author = {{Mart{\'\i}n-Navarro}, Ignacio and {Bakos}, Judit and {Trujillo}, Ignacio and {Knapen}, Johan H. and {Athanassoula}, E. and {Bosma}, Albert and {Comer{\'o}n}, S{\'e}bastien and {Elmegreen}, Bruce G. and {Erroz-Ferrer}, Santiago and {Gadotti}, Dimitri A. and {Gil de Paz}, Armando and {Hinz}, Joannah L. and {Ho}, Luis C. and {Holwerda}, Benne W. and {Kim}, Taehyun and {Laine}, Jarkko and {Laurikainen}, Eija and {Men{\'e}ndez-Delmestre}, Kar{\'\i}n. and {Mizusawa}, Trisha and {Mu{\~n}oz-Mateos}, Juan-Carlos and {Regan}, Michael W. and {Salo}, Heikki and {Seibert}, Mark and {Sheth}, Kartik},
        title = "{A unified picture of breaks and truncations in spiral galaxies from SDSS and S$^{4}$G imaging}",
      journal = {\mnras},
         year = 2012,
        month = dec,
       volume = {427},
       number = {2},
        pages = {1102-1134},
          doi = {10.1111/j.1365-2966.2012.21929.x},
archivePrefix = {arXiv},
       eprint = {1208.2893},
 primaryClass = {astro-ph.CO},
       adsurl = {https://ui.adsabs.harvard.edu/abs/2012MNRAS.427.1102M}
}

@ARTICLE{2006Debattista,
       author = {{Debattista}, Victor P. and {Mayer}, Lucio and {Carollo}, C. Marcella and {Moore}, Ben and {Wadsley}, James and {Quinn}, Thomas},
        title = "{The Secular Evolution of Disk Structural Parameters}",
      journal = {\apj},
         year = 2006,
        month = jul,
       volume = {645},
       number = {1},
        pages = {209-227},
          doi = {10.1086/504147},
archivePrefix = {arXiv},
       eprint = {astro-ph/0509310},
 primaryClass = {astro-ph},
       adsurl = {https://ui.adsabs.harvard.edu/abs/2006ApJ...645..209D}
}

@ARTICLE{2008Roskar,
       author = {{Ro{\v{s}}kar}, Rok and {Debattista}, Victor P. and {Stinson}, Gregory S. and {Quinn}, Thomas R. and {Kaufmann}, Tobias and {Wadsley}, James},
        title = "{Beyond Inside-Out Growth: Formation and Evolution of Disk Outskirts}",
      journal = {\apjl},
         year = 2008,
        month = mar,
       volume = {675},
       number = {2},
        pages = {L65},
          doi = {10.1086/586734},
archivePrefix = {arXiv},
       eprint = {0710.5523},
 primaryClass = {astro-ph},
       adsurl = {https://ui.adsabs.harvard.edu/abs/2008ApJ...675L..65R}
}

@ARTICLE{2008Bakos,
       author = {{Bakos}, Judit and {Trujillo}, Ignacio and {Pohlen}, Michael},
        title = "{Color Profiles of Spiral Galaxies: Clues on Outer-Disk Formation Scenarios}",
      journal = {\apjl},
         year = 2008,
        month = aug,
       volume = {683},
       number = {2},
        pages = {L103},
          doi = {10.1086/591671},
archivePrefix = {arXiv},
       eprint = {0807.2776},
 primaryClass = {astro-ph},
       adsurl = {https://ui.adsabs.harvard.edu/abs/2008ApJ...683L.103B}
}

@misc{photutils,
  author       = {Larry Bradley and
                  Brigitta Sipőcz and
                  Thomas Robitaille and
                  Erik Tollerud and
                  Zé Vinícius and
                  Christoph Deil and
                  Kyle Barbary and
                  Tom J Wilson and
                  Ivo Busko and
                  Axel Donath and
                  Hans Moritz Günther and
                  Mihai Cara and
                  P. L. Lim and
                  Sebastian Meßlinger and
                  Zach Burnett and
                  Simon Conseil and
                  Michael Droettboom and
                  Azalee Bostroem and
                  E. M. Bray and
                  Lars Andersen Bratholm and
                  William Jamieson and
                  Adam Ginsburg and
                  Geert Barentsen and
                  Matt Craig and
                  Sergio Pascual and
                  Shivangee Rathi and
                  Marshall Perrin and
                  Brett M. Morris},
  title        = {astropy/photutils: 2.2.0},
  month        = feb,
  year         = 2025,
  publisher    = {Zenodo},
  version      = {2.2.0},
  doi          = {10.5281/zenodo.14889440},
  url          = {https://doi.org/10.5281/zenodo.14889440},
  swhid        = {swh:1:dir:11159107f27a28985192ed1118b1f2055709d093
                   ;origin=https://doi.org/10.5281/zenodo.596036;visi
                   t=swh:1:snp:ae8c4a55d349d43e53cfe9ce92e678fcfe840f
                   3b;anchor=swh:1:rel:0117f67e8888adcdfc85308287dd9c
                   854b466389;path=astropy-photutils-ffb96c5
                  },
}

@ARTICLE{1979van_der_Kruit,
       author = {{van der Kruit}, P.~C.},
        title = "{Optical surface photometry of eight spiral galaxies studied in Westerbork.}",
      journal = {\aaps},
         year = 1979,
        month = oct,
       volume = {38},
        pages = {15-38},
       adsurl = {https://ui.adsabs.harvard.edu/abs/1979A&AS...38...15V}
}

@ARTICLE{2002Pohlen,
       author = {{Pohlen}, M. and {Dettmar}, R.-J. and {L{\"u}tticke}, R. and {Aronica}, G.},
        title = "{Outer edges of face-on spiral galaxies. Deep optical imaging of NGC 5923, UGC 9837 and NGC 5434}",
      journal = {\aap},
         year = 2002,
        month = sep,
       volume = {392},
        pages = {807-816},
          doi = {10.1051/0004-6361:20020994},
       adsurl = {https://ui.adsabs.harvard.edu/abs/2002A&A...392..807P}
}

@ARTICLE{2005Erwin,
       author = {{Erwin}, Peter and {Beckman}, John E. and {Pohlen}, Michael},
        title = "{Antitruncation of Disks in Early-Type Barred Galaxies}",
      journal = {\apjl},
         year = 2005,
        month = jun,
       volume = {626},
       number = {2},
        pages = {L81-L84},
          doi = {10.1086/431739},
archivePrefix = {arXiv},
       eprint = {astro-ph/0505216},
 primaryClass = {astro-ph},
       adsurl = {https://ui.adsabs.harvard.edu/abs/2005ApJ...626L..81E}
}

@ARTICLE{2008Erwin,
       author = {{Erwin}, Peter and {Pohlen}, Michael and {Beckman}, John E.},
        title = "{The Outer Disks of Early-Type Galaxies. I. Surface-Brightness Profiles of Barred Galaxies}",
      journal = {\aj},
         year = 2008,
        month = jan,
       volume = {135},
       number = {1},
        pages = {20-54},
          doi = {10.1088/0004-6256/135/1/20},
archivePrefix = {arXiv},
       eprint = {0709.3505},
 primaryClass = {astro-ph},
       adsurl = {https://ui.adsabs.harvard.edu/abs/2008AJ....135...20E}
}

@ARTICLE{2001Ferguson,
       author = {{Ferguson}, A.~M.~N. and {Clarke}, C.~J.},
        title = "{The evolution of stellar exponential discs}",
      journal = {\mnras},
         year = 2001,
        month = aug,
       volume = {325},
       number = {2},
        pages = {781-791},
          doi = {10.1046/j.1365-8711.2001.04501.x},
archivePrefix = {arXiv},
       eprint = {astro-ph/0103205},
 primaryClass = {astro-ph},
       adsurl = {https://ui.adsabs.harvard.edu/abs/2001MNRAS.325..781F}
}

@ARTICLE{1987van_der_Kruit,
       author = {{van der Kruit}, P.~C.},
        title = "{The radial distribution of surface brightness in galactic disks.}",
      journal = {\aap},
         year = 1987,
        month = feb,
       volume = {173},
        pages = {59-80},
       adsurl = {https://ui.adsabs.harvard.edu/abs/1987A&A...173...59V}
}

@ARTICLE{2020Tang,
       author = {{Tang}, Yimeng and {Chen}, Qianhui and {Zhang}, Hong-Xin and {Lin}, Zesen and {Chen}, Guangwen and {Gao}, Yulong and {Liang}, Zhixiong and {Liu}, Haiyang and {Kong}, Xu},
        title = "{New Constraints on the Origin of Surface Brightness Profile Breaks of Disk Galaxies from MaNGA}",
      journal = {\apj},
         year = 2020,
        month = jul,
       volume = {897},
       number = {1},
          eid = {79},
        pages = {79},
          doi = {10.3847/1538-4357/ab98fd},
archivePrefix = {arXiv},
       eprint = {2006.01356},
 primaryClass = {astro-ph.GA},
       adsurl = {https://ui.adsabs.harvard.edu/abs/2020ApJ...897...79T}
}

@ARTICLE{Guti2011,
       author = {{Guti{\'e}rrez}, Leonel and {Erwin}, Peter and {Aladro}, Rebeca and {Beckman}, John E.},
        title = "{The Outer Disks of Early-type Galaxies. II. Surface-brightness Profiles of Unbarred Galaxies and Trends with Hubble Type}",
      journal = {\aj},
         year = 2011,
        month = nov,
       volume = {142},
       number = {5},
          eid = {145},
        pages = {145},
          doi = {10.1088/0004-6256/142/5/145},
archivePrefix = {arXiv},
       eprint = {1108.3662},
 primaryClass = {astro-ph.CO},
       adsurl = {https://ui.adsabs.harvard.edu/abs/2011AJ....142..145G}
}

@ARTICLE{Marino2016,
       author = {{Marino}, R.~A. and {Gil de Paz}, A. and {S{\'a}nchez}, S.~F. and {S{\'a}nchez-Bl{\'a}zquez}, P. and {Cardiel}, N. and {Castillo-Morales}, A. and {Pascual}, S. and {V{\'\i}lchez}, J. and {Kehrig}, C. and {Moll{\'a}}, M. and {Mendez-Abreu}, J. and {Catal{\'a}n-Torrecilla}, C. and {Florido}, E. and {Perez}, I. and {Ruiz-Lara}, T. and {Ellis}, S. and {L{\'o}pez-S{\'a}nchez}, A.~R. and {Gonz{\'a}lez Delgado}, R.~M. and {de Lorenzo-C{\'a}ceres}, A. and {Garc{\'\i}a-Benito}, R. and {Galbany}, L. and {Zibetti}, S. and {Cortijo}, C. and {Kalinova}, V. and {Mast}, D. and {Iglesias-P{\'a}ramo}, J. and {Papaderos}, P. and {Walcher}, C.~J. and {Bland-Hawthorn}, J.},
        title = "{Outer-disk reddening and gas-phase metallicities: The CALIFA connection}",
      journal = {\aap},
         year = 2016,
        month = jan,
       volume = {585},
          eid = {A47},
        pages = {A47},
          doi = {10.1051/0004-6361/201526986},
archivePrefix = {arXiv},
       eprint = {1509.07878},
 primaryClass = {astro-ph.GA},
       adsurl = {https://ui.adsabs.harvard.edu/abs/2016A&A...585A..47M}
}

@ARTICLE{2009_James,
       author = {{James}, P.~A. and {Bretherton}, C.~F. and {Knapen}, J.~H.},
        title = "{The H{\ensuremath{\alpha}} galaxy survey. VII. The spatial distribution of star formation within disks and bulges}",
      journal = {\aap},
         year = 2009,
        month = jul,
       volume = {501},
       number = {1},
        pages = {207-220},
          doi = {10.1051/0004-6361/200810715},
archivePrefix = {arXiv},
       eprint = {0904.4261},
 primaryClass = {astro-ph.CO},
       adsurl = {https://ui.adsabs.harvard.edu/abs/2009A&A...501..207J}
}

@ARTICLE{2020a_Nersesian,
       author = {{Nersesian}, Angelos and {Verstocken}, Sam and {Viaene}, S{\'e}bastien and {Baes}, Maarten and {Xilouris}, Emmanuel M. and {Bianchi}, Simone and {Casasola}, Viviana and {Clark}, Christopher J.~R. and {Davies}, Jonathan I. and {De Looze}, Ilse and {De Vis}, Pieter and {Dobbels}, Wouter and {Fritz}, Jacopo and {Galametz}, Maud and {Galliano}, Fr{\'e}d{\'e}ric and {Jones}, Anthony P. and {Madden}, Suzanne C. and {Mosenkov}, Aleksandr V. and {Tr{\v{c}}ka}, Ana and {Ysard}, Nathalie},
        title = "{High-resolution, 3D radiative transfer modelling. III. The DustPedia barred galaxies}",
      journal = {\aap},
         year = 2020,
        month = may,
       volume = {637},
          eid = {A25},
        pages = {A25},
          doi = {10.1051/0004-6361/201936176},
archivePrefix = {arXiv},
       eprint = {2004.03616},
 primaryClass = {astro-ph.GA},
       adsurl = {https://ui.adsabs.harvard.edu/abs/2020A&A...637A..25N}
}

@ARTICLE{2007_Wozniak,
       author = {{Wozniak}, H.},
        title = "{The distribution of stellar population age in galactic bars}",
      journal = {\aap},
         year = 2007,
        month = apr,
       volume = {465},
       number = {1},
        pages = {L1-L4},
          doi = {10.1051/0004-6361:20067020},
archivePrefix = {arXiv},
       eprint = {astro-ph/0701687},
 primaryClass = {astro-ph},
       adsurl = {https://ui.adsabs.harvard.edu/abs/2007A&A...465L...1W}
}

@ARTICLE{2020_Fraser,
       author = {{Fraser-McKelvie}, Amelia and {Arag{\'o}n-Salamanca}, Alfonso and {Merrifield}, Michael and {Masters}, Karen and {Nair}, Preethi and {Emsellem}, Eric and {Kraljic}, Katarina and {Krishnarao}, Dhanesh and {Andrews}, Brett H. and {Drory}, Niv and {Neumann}, Justus},
        title = "{SDSS-IV MaNGA: spatially resolved star formation in barred galaxies}",
      journal = {\mnras},
         year = 2020,
        month = jul,
       volume = {495},
       number = {4},
        pages = {4158-4169},
          doi = {10.1093/mnras/staa1416},
archivePrefix = {arXiv},
       eprint = {2005.08987},
 primaryClass = {astro-ph.GA},
       adsurl = {https://ui.adsabs.harvard.edu/abs/2020MNRAS.495.4158F}
}

@ARTICLE{galaxy_zoo,
       author = {{Walmsley}, Mike and {G{\'e}ron}, Tobias and {Kruk}, Sandor and {Scaife}, Anna M.~M. and {Lintott}, Chris and {Masters}, Karen L. and {Dawson}, James M. and {Dickinson}, Hugh and {Fortson}, Lucy and {Garland}, Izzy L. and {Mantha}, Kameswara and {O'Ryan}, David and {Popp}, J{\"u}rgen and {Simmons}, Brooke and {Baeten}, Elisabeth M. and {Macmillan}, Christine},
        title = "{Galaxy Zoo DESI: Detailed morphology measurements for 8.7M galaxies in the DESI Legacy Imaging Surveys}",
      journal = {\mnras},
         year = 2023,
        month = dec,
       volume = {526},
       number = {3},
        pages = {4768-4786},
          doi = {10.1093/mnras/stad2919},
archivePrefix = {arXiv},
       eprint = {2309.11425},
 primaryClass = {astro-ph.GA},
       adsurl = {https://ui.adsabs.harvard.edu/abs/2023MNRAS.526.4768W}
}

@ARTICLE{sed_1,
       author = {{Euclid Collaboration} and {Nersesian}, A. and {Abdurro'uf} and {Baes}, M. and {Tortora}, C. and {Kova{\v{c}}i{\'c}}, I. and {Bisigello}, L. and {Corcho-Caballero}, P. and {Dur{\'a}n-Camacho}, E. and {Hunt}, L.~K. and {Iglesias-Navarro}, P. and {Ragusa}, R. and {Rom{\'a}n}, J. and {Shankar}, F. and {Siudek}, M. and {Sorce}, J.~G. and {Marleau}, F.~R. and {Aghanim}, N. and {Andreon}, S. and {Auricchio}, N. and {Baccigalupi}, C. and {Baldi}, M. and {Bardelli}, S. and {Biviano}, A. and {Branchini}, E. and {Brescia}, M. and {Camera}, S. and {Ca{\~n}as-Herrera}, G. and {Capobianco}, V. and {Carbone}, C. and {Carretero}, J. and {Casas}, S. and {Castellano}, M. and {Castignani}, G. and {Cavuoti}, S. and {Cimatti}, A. and {Colodro-Conde}, C. and {Congedo}, G. and {Conselice}, C.~J. and {Conversi}, L. and {Copin}, Y. and {Courbin}, F. and {Courtois}, H.~M. and {Da Silva}, A. and {Degaudenzi}, H. and {De Lucia}, G. and {Dole}, H. and {Douspis}, M. and {Dubath}, F. and {Dupac}, X. and {Dusini}, S. and {Farina}, M. and {Farinelli}, R. and {Faustini}, F. and {Ferriol}, S. and {Finelli}, F. and {Fourmanoit}, N. and {Frailis}, M. and {Franceschi}, E. and {Fumana}, M. and {Galeotta}, S. and {George}, K. and {Gillis}, B. and {Giocoli}, C. and {Gracia-Carpio}, J. and {Grazian}, A. and {Grupp}, F. and {Haugan}, S.~V.~H. and {Holmes}, W. and {Hormuth}, F. and {Hornstrup}, A. and {Jahnke}, K. and {Jhabvala}, M. and {Keih{\"a}nen}, E. and {Kermiche}, S. and {Kilbinger}, M. and {Kubik}, B. and {K{\"u}mmel}, M. and {Kunz}, M. and {Kurki-Suonio}, H. and {Le Brun}, A.~M.~C. and {Ligori}, S. and {Lilje}, P.~B. and {Lindholm}, V. and {Lloro}, I. and {Mainetti}, G. and {Maino}, D. and {Maiorano}, E. and {Mansutti}, O. and {Marggraf}, O. and {Martinelli}, M. and {Martinet}, N. and {Marulli}, F. and {Massey}, R.~J. and {Medinaceli}, E. and {Mei}, S. and {Melchior}, M. and {Mellier}, Y. and {Meneghetti}, M. and {Merlin}, E. and {Meylan}, G. and {Mora}, A. and {Moresco}, M. and {Moscardini}, L. and {Neissner}, C. and {Niemi}, S.~M. and {Padilla}, C. and {Paltani}, S. and {Pasian}, F. and {Pedersen}, K. and {Pettorino}, V. and {Pires}, S. and {Polenta}, G. and {Poncet}, M. and {Popa}, L.~A. and {Pozzetti}, L. and {Renzi}, A. and {Rhodes}, J. and {Riccio}, G. and {Romelli}, E. and {Roncarelli}, M. and {Saglia}, R. and {Sakr}, Z. and {Sapone}, D. and {Sartoris}, B. and {Schneider}, P. and {Schrabback}, T. and {Secroun}, A. and {Seidel}, G. and {Seiffert}, M. and {Serrano}, S. and {Simon}, P. and {Sirignano}, C. and {Sirri}, G. and {Stanco}, L. and {Steinwagner}, J. and {Tallada-Cresp{\'\i}}, P. and {Taylor}, A.~N. and {Tereno}, I. and {Toft}, S. and {Toledo-Moreo}, R. and {Torradeflot}, F. and {Tutusaus}, I. and {Valenziano}, L. and {Valiviita}, J. and {Vassallo}, T. and {Verdoes Kleijn}, G. and {Veropalumbo}, A. and {Wang}, Y. and {Weller}, J. and {Zamorani}, G. and {Zerbi}, F.~M. and {Zinchenko}, I.~A. and {Zucca}, E. and {Allevato}, V. and {Bolzonella}, M. and {Bozzo}, E. and {Burigana}, C. and {Cabanac}, R. and {Calabrese}, M. and {Cappi}, A. and {Escartin Vigo}, J.~A. and {Gabarra}, L. and {Mart{\'\i}n-Fleitas}, J. and {Matthew}, S. and {Mauri}, N. and {Metcalf}, R.~B. and {Nucita}, A.~A. and {Pezzotta}, A. and {P{\"o}ntinen}, M. and {Porciani}, C. and {Risso}, I. and {Scottez}, V. and {Sereno}, M. and {Tenti}, M. and {Viel}, M. and {Wiesmann}, M. and {Akrami}, Y. and {Andika}, I.~T. and {Anselmi}, S. and {Archidiacono}, M. and {Atrio-Barandela}, F. and {Bertacca}, D. and {Bethermin}, M. and {Blanchard}, A. and {Blot}, L. and {Borgani}, S. and {Brown}, M.~L. and {Bruton}, S. and {Calabro}, A. and {Camacho Quevedo}, B. and {Caro}, F. and {Carvalho}, C.~S. and {Castro}, T. and {Cogato}, F. and {Conseil}, S. and {Cooray}, A.~R. and {Cucciati}, O. and {Davini}, S. and {De Paolis}, F.},
        title = "{Euclid preparation: LXXXI. The impact of nonparametric star formation histories on spatially resolved galaxy property estimation using synthetic Euclid images}",
      journal = {\aap},
         year = 2026,
        month = mar,
       volume = {707},
          eid = {A231},
        pages = {A231},
          doi = {10.1051/0004-6361/202556464},
archivePrefix = {arXiv},
       eprint = {2511.22399},
 primaryClass = {astro-ph.GA},
       adsurl = {https://ui.adsabs.harvard.edu/abs/2026A&A...707A.231E}
}

@ARTICLE{sed_2,
       author = {{Euclid Collaboration} and {Abdurro'Uf} and {Tortora}, C. and {Baes}, M. and {Nersesian}, A. and {Kovacic}, I. and {Bolzonella}, M. and {Lancon}, A. and {Bisigello}, L. and {Annibali}, F. and {Bremer}, M.~N. and {Carollo}, D. and {Conselice}, C.~J. and {Enia}, A. and {Ferguson}, A.~M.~N. and {Ferre-Mateu}, A. and {Hunt}, L.~K. and {Iodice}, E. and {Knapen}, J.~H. and {Iovino}, A. and {Marleau}, F.~R. and {Peletier}, R.~F. and {Ragusa}, R. and {Rejkuba}, M. and {Robotham}, A.~S.~G. and {Roman}, J. and {Saifollahi}, T. and {Salucci}, P. and {Scodeggio}, M. and {Siudek}, M. and {van der Wel}, A. and {Voggel}, K. and {Altieri}, B. and {Andreon}, S. and {Baccigalupi}, C. and {Baldi}, M. and {Bardelli}, S. and {Biviano}, A. and {Bonchi}, A. and {Bonino}, D. and {Branchini}, E. and {Brescia}, M. and {Brinchmann}, J. and {Caillat}, A. and {Camera}, S. and {Canas-Herrera}, G. and {Capobianco}, V. and {Carbone}, C. and {Carretero}, J. and {Casas}, S. and {Castellano}, M. and {Castignani}, G. and {Cavuoti}, S. and {Chambers}, K.~C. and {Cimatti}, A. and {Colodro-Conde}, C. and {Congedo}, G. and {Conversi}, L. and {Copin}, Y. and {Courbin}, F. and {Courtois}, H.~M. and {Cropper}, M. and {da Silva}, A. and {Degaudenzi}, H. and {de}, Lucia G. and {di}, Giorgio A.~M. and {Dinis}, J. and {Dole}, H. and {Dubath}, F. and {Dupac}, X. and {Dusini}, S. and {Escoffier}, S. and {Farina}, M. and {Farinelli}, R. and {Farrens}, S. and {Faustini}, F. and {Ferriol}, S. and {Finelli}, F. and {Fotopoulou}, S. and {Frailis}, M. and {Franceschi}, E. and {Fumana}, M. and {Galeotta}, S. and {Gillis}, B. and {Giocoli}, C. and {Gomez-Alvarez}, P. and {Gracia-Carpio}, J. and {Grazian}, A. and {Grupp}, F. and {Holmes}, W. and {Hormuth}, F. and {Hornstrup}, A. and {Hudelot}, P. and {Jahnke}, K. and {Jhabvala}, M. and {Keihanen}, E. and {Kermiche}, S. and {Kiessling}, A. and {Kilbinger}, M. and {Kubik}, B. and {Kummel}, M. and {Kunz}, M. and {Kurki-Suonio}, H. and {Le Brun}, A.~M.~C. and {Ligori}, S. and {Lilje}, P.~B. and {Lindholm}, V. and {Lloro}, I. and {Mainetti}, G. and {Maino}, D. and {Maiorano}, E. and {Mansutti}, O. and {Marggraf}, O. and {Markovic}, K. and {Martinelli}, M. and {Martinet}, N. and {Marulli}, F. and {Massey}, R. and {Medinaceli}, E. and {Mei}, S. and {Melchior}, M. and {Mellier}, Y. and {Meneghetti}, M. and {Merlin}, E. and {Meylan}, G. and {Mora}, A. and {Moresco}, M. and {Moscardini}, L. and {Niemi}, S.-M. and {Nightingale}, J.~W. and {Padilla}, C. and {Paltani}, S. and {Pasian}, F. and {Pedersen}, K. and {Pettorino}, V. and {Polenta}, G. and {Poncet}, M. and {Popa}, L.~A. and {Pozzetti}, L. and {Raison}, F. and {Renzi}, A. and {Rhodes}, J. and {Riccio}, G. and {Romelli}, E. and {Roncarelli}, M. and {Rossetti}, E. and {Saglia}, R. and {Sakr}, Z. and {Sapone}, D. and {Sartoris}, B. and {Schirmer}, M. and {Schneider}, P. and {Schrabback}, T. and {Secroun}, A. and {Sefusatti}, E. and {Seidel}, G. and {Serrano}, S. and {Simon}, P. and {Sirignano}, C. and {Sirri}, G. and {Stanco}, L. and {Steinwagner}, J. and {Tallada-Crespi}, P. and {Taylor}, A.~N. and {Tereno}, I. and {Toft}, S. and {Toledo-Moreo}, R. and {Torradeflot}, F. and {Tutusaus}, I. and {Valenziano}, L. and {Valiviita}, J. and {Vassallo}, T. and {Verdoes Kleijn}, G. and {Veropalumbo}, A. and {Wang}, Y. and {Weller}, J. and {Zamorani}, G. and {Zucca}, E. and {Bozzo}, E. and {Burigana}, C. and {Calabrese}, M. and {di Ferdinando}, D. and {Escartin Vigo}, J.~A. and {Matthew}, S. and {Mauri}, N. and {Pontinen}, M. and {Porciani}, C. and {Scottez}, V. and {Tenti}, M. and {Viel}, M. and {Wiesmann}, M. and {Akrami}, Y. and {Allevato}, V. and {Anselmi}, S. and {Archidiacono}, M. and {Atrio-Barandela}, F. and {Ballardini}, M. and {Bertacca}, D. and {Blanchard}, A. and {Blot}, L.},
        title = "{Euclid preparation: LXXIII. Spatially resolved stellar populations of local galaxies with Euclid: A proof of concept using synthetic images with the TNG50 simulation}",
      journal = {\aap},
         year = 2025,
        month = oct,
       volume = {702},
          eid = {A72},
        pages = {A72},
          doi = {10.1051/0004-6361/202554516},
archivePrefix = {arXiv},
       eprint = {2503.15635},
 primaryClass = {astro-ph.GA},
       adsurl = {https://ui.adsabs.harvard.edu/abs/2025A&A...702A..72E}
}

@ARTICLE{sed_3,
       author = {{Euclid Collaboration} and {Kovacic}, I. and {Baes}, M. and {Nersesian}, A. and {Andreadis}, N. and {Nemani}, L. and {Abdurro'Uf} and {Bisigello}, L. and {Bolzonella}, M. and {Tortora}, C. and {van der Wel}, A. and {Cavuoti}, S. and {Conselice}, C.~J. and {Enia}, A. and {Hunt}, L.~K. and {Iglesias-Navarro}, P. and {Iodice}, E. and {Knapen}, J.~H. and {Marleau}, F.~R. and {Muller}, O. and {Peletier}, R.~F. and {Roman}, J. and {Ragusa}, R. and {Salucci}, P. and {Saifollahi}, T. and {Scodeggio}, M. and {Siudek}, M. and {de Waele}, T. and {Amara}, A. and {Andreon}, S. and {Auricchio}, N. and {Baccigalupi}, C. and {Baldi}, M. and {Bardelli}, S. and {Battaglia}, P. and {Bender}, R. and {Bodendorf}, C. and {Bonino}, D. and {Bon}, W. and {Branchini}, E. and {Brescia}, M. and {Brinchmann}, J. and {Camera}, S. and {Capobianco}, V. and {Carbone}, C. and {Carretero}, J. and {Casas}, S. and {Castander}, F.~J. and {Castellano}, M. and {Castignani}, G. and {Cimatti}, A. and {Colodro-Conde}, C. and {Congedo}, G. and {Conversi}, L. and {Copin}, Y. and {Courbin}, F. and {Courtois}, H.~M. and {da Silva}, A. and {Degaudenzi}, H. and {de}, Lucia G. and {di}, Giorgio A.~M. and {Dinis}, J. and {Douspis}, M. and {Dubath}, F. and {Dupac}, X. and {Dusini}, S. and {Ealet}, A. and {Farina}, M. and {Farrens}, S. and {Faustini}, F. and {Ferriol}, S. and {Fosalba}, P. and {Frailis}, M. and {Franceschi}, E. and {Galeotta}, S. and {Gillis}, B. and {Giocoli}, C. and {Grazian}, A. and {Grupp}, F. and {Guzzo}, L. and {Haugan}, S.~V.~H. and {Holmes}, W. and {Hook}, I. and {Hormuth}, F. and {Hornstrup}, A. and {Jahnke}, K. and {Jhabvala}, M. and {Joachimi}, B. and {Keihanen}, E. and {Kermiche}, S. and {Kiessling}, A. and {Kilbinger}, M. and {Kubik}, B. and {Kuijken}, K. and {Kummel}, M. and {Kunz}, M. and {Kurki-Suonio}, H. and {Ligori}, S. and {Lilje}, P.~B. and {Lindholm}, V. and {Lloro}, I. and {Maino}, D. and {Maiorano}, E. and {Mansutti}, O. and {Marcin}, S. and {Marggraf}, O. and {Markovic}, K. and {Martinelli}, M. and {Martinet}, N. and {Marulli}, F. and {Massey}, R. and {Medinaceli}, E. and {Mei}, S. and {Melchior}, M. and {Mellier}, Y. and {Meneghetti}, M. and {Merlin}, E. and {Meylan}, G. and {Moresco}, M. and {Moscardini}, L. and {Niemi}, S.-M. and {Nightingale}, J.~W. and {Padilla}, C. and {Paltani}, S. and {Pasian}, F. and {Pedersen}, K. and {Pettorino}, V. and {Pires}, S. and {Polenta}, G. and {Poncet}, M. and {Popa}, L.~A. and {Pozzetti}, L. and {Raison}, F. and {Rebolo}, R. and {Renzi}, A. and {Rhodes}, J. and {Riccio}, G. and {Romelli}, E. and {Roncarelli}, M. and {Rossetti}, E. and {Saglia}, R. and {Sakr}, Z. and {Sanchez}, A.~G. and {Sapone}, D. and {Sartoris}, B. and {Schirmer}, M. and {Schneider}, P. and {Schrabback}, T. and {Secroun}, A. and {Seidel}, G. and {Serrano}, S. and {Sirignano}, C. and {Sirri}, G. and {Stanco}, L. and {Steinwagner}, J. and {Tallada-Crespi}, P. and {Tavagnacco}, D. and {Taylor}, A.~N. and {Teplitz}, H.~I. and {Tereno}, I. and {Toledo-Moreo}, R. and {Torradeflot}, F. and {Tutusaus}, I. and {Valenziano}, L. and {Vassallo}, T. and {Verdoes Kleijn}, G. and {Veropalumbo}, A. and {Wang}, Y. and {Weller}, J. and {Zacchei}, A. and {Zamorani}, G. and {Zucca}, E. and {Biviano}, A. and {Bozzo}, E. and {Burigana}, C. and {Calabrese}, M. and {di Ferdinando}, D. and {Escartin Vigo}, J.~A. and {Finelli}, F. and {Gracia-Carpio}, J. and {Matthew}, S. and {Mauri}, N. and {Pontinen}, M. and {Scottez}, V. and {Tenti}, M. and {Viel}, M. and {Wiesmann}, M. and {Akrami}, Y. and {Allevato}, V. and {Alvi}, S. and {Anselmi}, S. and {Archidiacono}, M. and {Atrio-Barandela}, F. and {Ballardini}, M. and {Bethermin}, M. and {Blot}, L. and {Borgani}, S. and {Bruton}, S. and {Cabanac}, R. and {Calabro}, A.},
        title = "{Euclid preparation: LXVIII. Extracting physical parameters from galaxies with machine learning}",
      journal = {\aap},
         year = 2025,
        month = mar,
       volume = {695},
          eid = {A284},
        pages = {A284},
          doi = {10.1051/0004-6361/202453111},
archivePrefix = {arXiv},
       eprint = {2501.14408},
 primaryClass = {astro-ph.GA},
       adsurl = {https://ui.adsabs.harvard.edu/abs/2025A&A...695A.284E}
}

@article{Kroupa2001,
       author = {{Kroupa}, Pavel},
        title = "{On the variation of the initial mass function}",
      journal = {\mnras},
         year = 2001,
        month = apr,
       volume = {322},
       number = {2},
        pages = {231-246},
          doi = {10.1046/j.1365-8711.2001.04022.x},
archivePrefix = {arXiv},
       eprint = {astro-ph/0009005},
 primaryClass = {astro-ph},
       adsurl = {https://ui.adsabs.harvard.edu/abs/2001MNRAS.322..231K}
}

@article{Bressan2012,
       author = {{Bressan}, Alessandro and {Marigo}, Paola and {Girardi}, L{\'e}o. and {Salasnich}, Bernardo and {Dal Cero}, Claudia and {Rubele}, Stefano and {Nanni}, Ambra},
        title = "{PARSEC: stellar tracks and isochrones with the PAdova and TRieste Stellar Evolution Code}",
      journal = {\mnras},
         year = 2012,
        month = nov,
       volume = {427},
       number = {1},
        pages = {127-145},
          doi = {10.1111/j.1365-2966.2012.21948.x},
archivePrefix = {arXiv},
       eprint = {1208.4498},
 primaryClass = {astro-ph.SR},
       adsurl = {https://ui.adsabs.harvard.edu/abs/2012MNRAS.427..127B}
}

@ARTICLE{2011Coelho,
       author = {{Coelho}, P. and {Gadotti}, D.~A.},
        title = "{Bars Rejuvenating Bulges? Evidence from Stellar Population Analysis}",
      journal = {\apjl},
         year = 2011,
        month = dec,
       volume = {743},
       number = {1},
          eid = {L13},
        pages = {L13},
          doi = {10.1088/2041-8205/743/1/L13},
archivePrefix = {arXiv},
       eprint = {1111.1736},
 primaryClass = {astro-ph.CO},
       adsurl = {https://ui.adsabs.harvard.edu/abs/2011ApJ...743L..13C}
}

@ARTICLE{2019Chown,
       author = {{Chown}, Ryan and {Li}, Cheng and {Athanassoula}, E. and {Li}, Niu and {Wilson}, Christine D. and {Lin}, Lin and {Mo}, Houjun and {Parker}, Laura C. and {Xiao}, Ting},
        title = "{Linking bar- and interaction-driven molecular gas concentration with centrally enhanced star formation in EDGE-CALIFA galaxies}",
      journal = {\mnras},
         year = 2019,
        month = apr,
       volume = {484},
       number = {4},
        pages = {5192-5211},
          doi = {10.1093/mnras/stz349},
archivePrefix = {arXiv},
       eprint = {1810.08624},
 primaryClass = {astro-ph.GA},
       adsurl = {https://ui.adsabs.harvard.edu/abs/2019MNRAS.484.5192C}
}

@ARTICLE{2009Aguerri,
       author = {{Aguerri}, J.~A.~L. and {M{\'e}ndez-Abreu}, J. and {Corsini}, E.~M.},
        title = "{The population of barred galaxies in the local universe. I. Detection and characterisation of bars}",
      journal = {\aap},
         year = 2009,
        month = feb,
       volume = {495},
       number = {2},
        pages = {491-504},
          doi = {10.1051/0004-6361:200810931},
archivePrefix = {arXiv},
       eprint = {0901.2346},
 primaryClass = {astro-ph.GA},
       adsurl = {https://ui.adsabs.harvard.edu/abs/2009A&A...495..491A}
}

@ARTICLE{2015Willett,
       author = {{Willett}, Kyle W. and {Schawinski}, Kevin and {Simmons}, Brooke D. and {Masters}, Karen L. and {Skibba}, Ramin A. and {Kaviraj}, Sugata and {Melvin}, Thomas and {Wong}, O. Ivy and {Nichol}, Robert C. and {Cheung}, Edmond and {Lintott}, Chris J. and {Fortson}, Lucy},
        title = "{Galaxy Zoo: the dependence of the star formation-stellar mass relation on spiral disc morphology}",
      journal = {\mnras},
         year = 2015,
        month = may,
       volume = {449},
       number = {1},
        pages = {820-827},
          doi = {10.1093/mnras/stv307},
archivePrefix = {arXiv},
       eprint = {1502.03444},
 primaryClass = {astro-ph.GA},
       adsurl = {https://ui.adsabs.harvard.edu/abs/2015MNRAS.449..820W}
}

@ARTICLE{2016James,
       author = {{James}, P.~A. and {Percival}, S.~M.},
        title = "{Stellar population constraints on the ages of galactic bars}",
      journal = {\mnras},
         year = 2016,
        month = mar,
       volume = {457},
       number = {1},
        pages = {917-925},
          doi = {10.1093/mnras/stv2978},
archivePrefix = {arXiv},
       eprint = {1601.04706},
 primaryClass = {astro-ph.GA},
       adsurl = {https://ui.adsabs.harvard.edu/abs/2016MNRAS.457..917J}
}

@ARTICLE{2018Khoperskov,
       author = {{Khoperskov}, S. and {Haywood}, M. and {Di Matteo}, P. and {Lehnert}, M.~D. and {Combes}, F.},
        title = "{Bar quenching in gas-rich galaxies}",
      journal = {\aap},
         year = 2018,
        month = jan,
       volume = {609},
          eid = {A60},
        pages = {A60},
          doi = {10.1051/0004-6361/201731211},
archivePrefix = {arXiv},
       eprint = {1709.03604},
 primaryClass = {astro-ph.GA},
       adsurl = {https://ui.adsabs.harvard.edu/abs/2018A&A...609A..60K}
}

@ARTICLE{2020Newnham,
       author = {{Newnham}, L. and {Hess}, Kelley M. and {Masters}, Karen L. and {Kruk}, Sandor and {Penny}, Samantha J. and {Lingard}, Tim and {Smethurst}, R.~J.},
        title = "{The H I morphology and stellar properties of strongly barred galaxies: support for bar quenching in massive spirals}",
      journal = {\mnras},
         year = 2020,
        month = mar,
       volume = {492},
       number = {4},
        pages = {4697-4715},
          doi = {10.1093/mnras/staa064},
archivePrefix = {arXiv},
       eprint = {1901.01164},
 primaryClass = {astro-ph.GA},
       adsurl = {https://ui.adsabs.harvard.edu/abs/2020MNRAS.492.4697N}
}

@INPROCEEDINGS{2012Oh,
       author = {{Oh}, Seulhee and {Oh}, K. and {Yi}, S.~K.},
        title = "{The Role of Large-scale Bars on Central Star Formation and AGN Activity}",
    booktitle = {American Astronomical Society Meeting Abstracts \#219},
         year = 2012,
       series = {American Astronomical Society Meeting Abstracts},
       volume = {219},
        month = jan,
          eid = {323.01},
        pages = {323.01},
       adsurl = {https://ui.adsabs.harvard.edu/abs/2012AAS...21932301O}
}

@ARTICLE{2012Masters,
       author = {{Masters}, Karen L. and {Nichol}, Robert C. and {Haynes}, Martha P. and {Keel}, William C. and {Lintott}, Chris and {Simmons}, Brooke and {Skibba}, Ramin and {Bamford}, Steven and {Giovanelli}, Riccardo and {Schawinski}, Kevin},
        title = "{Galaxy Zoo and ALFALFA: atomic gas and the regulation of star formation in barred disc galaxies}",
      journal = {\mnras},
         year = 2012,
        month = aug,
       volume = {424},
       number = {3},
        pages = {2180-2192},
          doi = {10.1111/j.1365-2966.2012.21377.x},
archivePrefix = {arXiv},
       eprint = {1205.5271},
 primaryClass = {astro-ph.CO},
       adsurl = {https://ui.adsabs.harvard.edu/abs/2012MNRAS.424.2180M}
}

@ARTICLE{2021Garcia,
       author = {{D{\'\i}az-Garc{\'\i}a}, S. and {Lisenfeld}, U. and {P{\'e}rez}, I. and {Zurita}, A. and {Verley}, S. and {Combes}, F. and {Espada}, D. and {Leon}, S. and {Mart{\'\i}nez-Badenes}, V. and {Sabater}, J. and {Verdes-Montenegro}, L.},
        title = "{Molecular gas and star formation within 12 strong galactic bars observed with IRAM-30 m}",
      journal = {\aap},
         year = 2021,
        month = oct,
       volume = {654},
          eid = {A135},
        pages = {A135},
          doi = {10.1051/0004-6361/202140674},
archivePrefix = {arXiv},
       eprint = {2106.13099},
 primaryClass = {astro-ph.GA},
       adsurl = {https://ui.adsabs.harvard.edu/abs/2021A&A...654A.135D}
}

@ARTICLE{2021Geron,
       author = {{G{\'e}ron}, Tobias and {Smethurst}, R.~J. and {Lintott}, Chris and {Kruk}, Sandor and {Masters}, Karen L. and {Simmons}, Brooke and {Stark}, David V.},
        title = "{Galaxy zoo: stronger bars facilitate quenching in star-forming galaxies}",
      journal = {\mnras},
         year = 2021,
        month = nov,
       volume = {507},
       number = {3},
        pages = {4389-4408},
          doi = {10.1093/mnras/stab2064},
archivePrefix = {arXiv},
       eprint = {2107.06913},
 primaryClass = {astro-ph.GA},
       adsurl = {https://ui.adsabs.harvard.edu/abs/2021MNRAS.507.4389G}
}

@ARTICLE{cardelli1989,
       author = {{Cardelli}, Jason A. and {Clayton}, Geoffrey C. and {Mathis}, John S.},
        title = "{The Relationship between Infrared, Optical, and Ultraviolet Extinction}",
      journal = {\apj},
         year = 1989,
        month = oct,
       volume = {345},
        pages = {245},
          doi = {10.1086/167900},
       adsurl = {https://ui.adsabs.harvard.edu/abs/1989ApJ...345..245C}
}

@INPROCEEDINGS{yuan2020,
       author = {{Yuan}, Xiangyan and {Li}, Zhengyang and {Liu}, Xiaowei and {Niu}, Dongsheng and {Lu}, Qishui and {Jiang}, Fanghua and {Wang}, Yuefei and {Li}, Xiaoyan and {Liang}, YongJun and {Wang}, Hai and {Zhang}, Chao and {Wang}, Jinfeng and {Li}, Bo and {Tian}, Jie and {Lu}, Haiping and {Chen}, Bingqiu and {Huang}, Yang and {Liu}, Xiangkun and {Yao}, Zhengqiu and {Cui}, Xiangqun and {Li}, Guoping},
        title = "{Development of the Multi-channel Photometric Survey telescope}",
    booktitle = {Ground-based and Airborne Telescopes VIII},
         year = 2020,
       editor = {{Marshall}, Heather K. and {Spyromilio}, Jason and {Usuda}, Tomonori},
       series = {Society of Photo-Optical Instrumentation Engineers (SPIE) Conference Series},
       volume = {11445},
        month = dec,
          eid = {114457M},
        pages = {114457M},
          doi = {10.1117/12.2562334},
       adsurl = {https://ui.adsabs.harvard.edu/abs/2020SPIE11445E..7MY}
}

@ARTICLE{2025Du,
       author = {{Du}, Guowang and {Cheng}, Yehao and {Yang}, Yuan-Pei and {Yang}, Jun and {Zhang}, Jinghua and {Zhu}, Dan and {Pan}, Yu and {Fang}, Yuan and {Zou}, Xingzhu and {Kumar}, Brajesh and {Guo}, Helong and {Zhu}, Xufeng and {Zhang}, Yangwei and {Kong}, Fanchuan and {Shang}, Chenxi and {Chen}, Xinlei and {Liu}, Xiangkun and {Liu}, Xiaowei},
        title = "{Multiwavelength Analysis of GRB 250101A: From Gamma-Ray Prompt Emission to Optical Afterglow}",
      journal = {\apj},
         year = 2025,
        month = aug,
       volume = {989},
       number = {1},
          eid = {19},
        pages = {19},
          doi = {10.3847/1538-4357/adea3d},
archivePrefix = {arXiv},
       eprint = {2503.15805},
 primaryClass = {astro-ph.HE},
       adsurl = {https://ui.adsabs.harvard.edu/abs/2025ApJ...989...19D}
}
\bibliographystyle{aasjournalv7}



\appendix

\renewcommand{\thefigure}{A\arabic{figure}}
\setcounter{figure}{0}

\begin{figure}[ht]
    \centering
    \includegraphics[width=\textwidth]{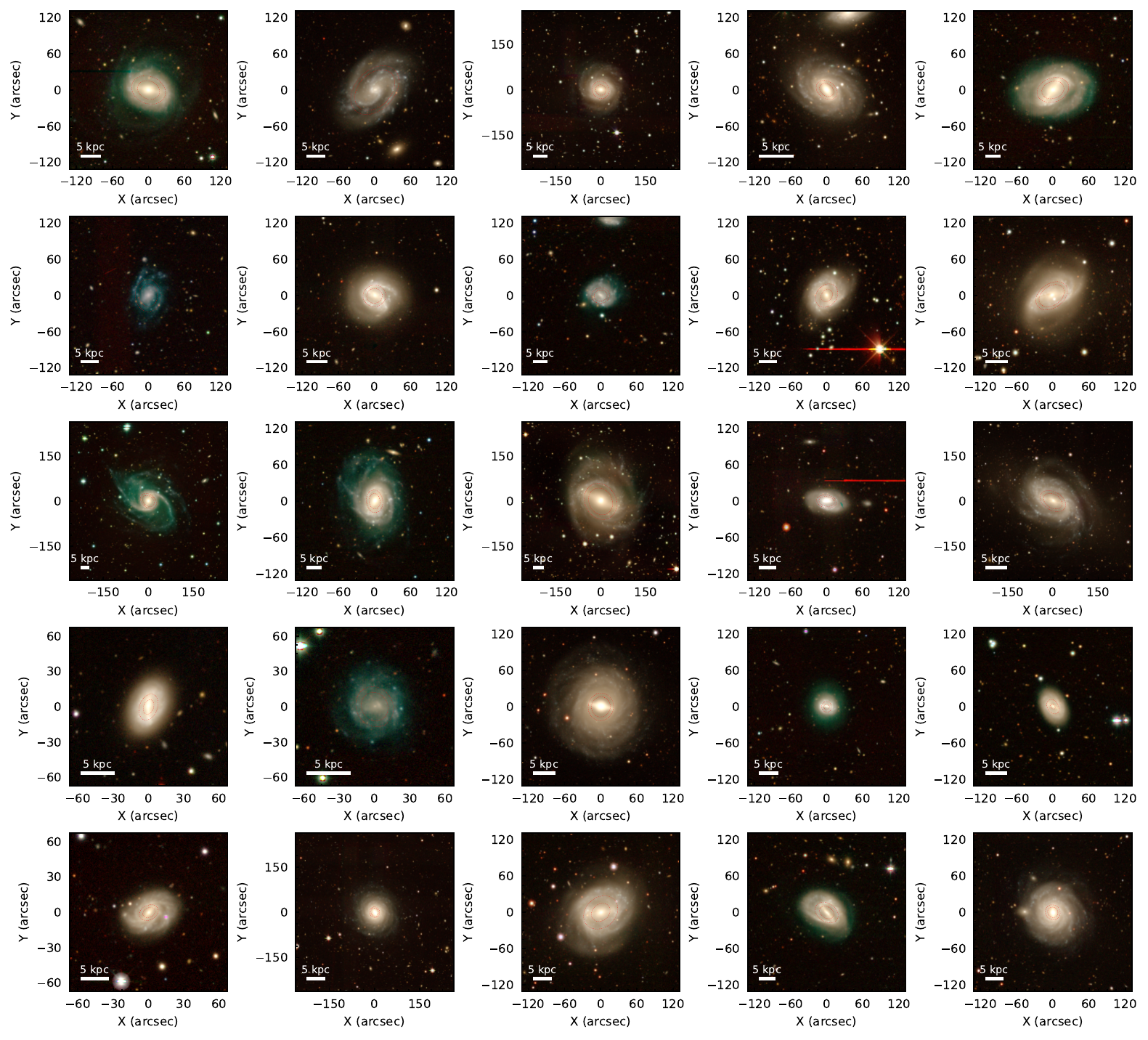}
    \caption{False-color $(g,r,z)$ composite images of barred galaxies in the MWL sample. The two red elliptical rings mark the respective plateau regions in each galaxy.}
    \label{fig:all_2}
\end{figure}

%\addtocounter{figure}{-1}

\begin{figure}[ht]
    \centering
    \includegraphics[width=\textwidth]{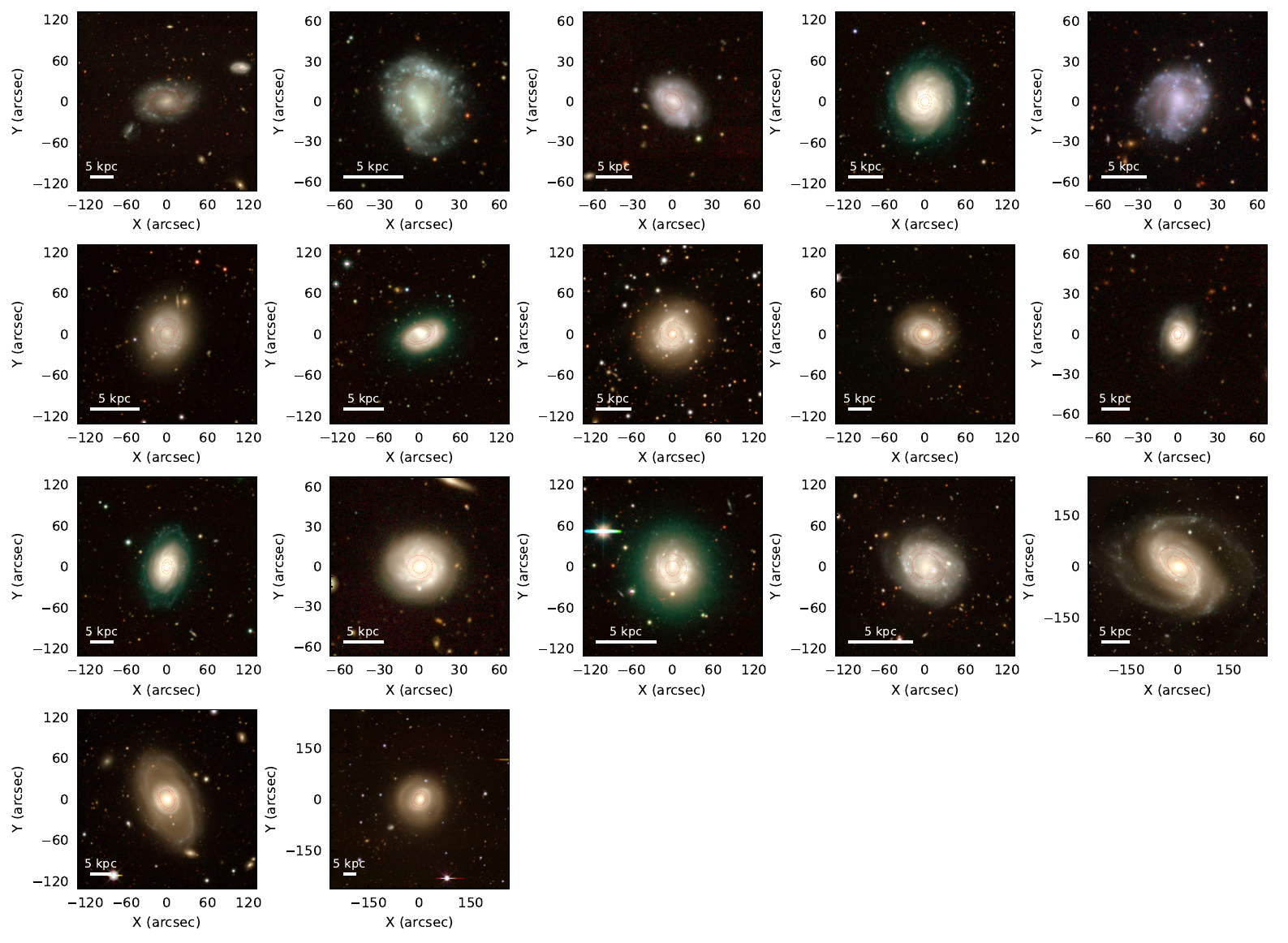}
    \caption{False-color $(g,r,z)$ composite images of non-barred galaxies in the MWL sample. The two red elliptical rings mark the respective plateau regions in each galaxy.}
    \label{fig:all_1}
\end{figure}

\begin{figure}[ht]
    \centering
    \includegraphics[width=\textwidth]{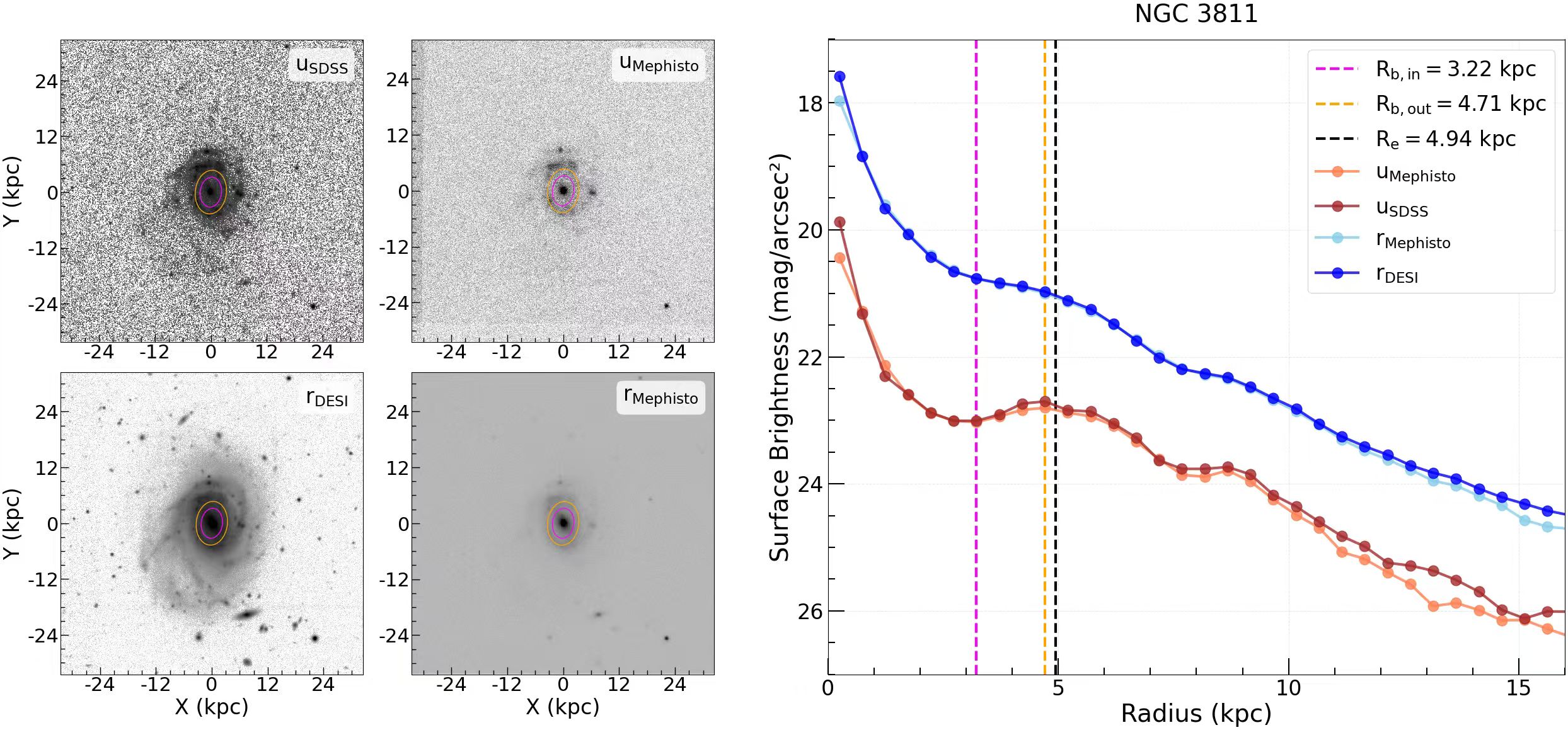}
    \caption{Multi-band images of NGC 3811 from Mephisto, SDSS, and DESI, and comparison of its radial surface brightness profiles derived from Mephisto, SDSS, and DESI imaging data. The plateau region is marked by red and yellow elliptical rings in the multi-band images.}
    \label{fig:diff_2}
\end{figure}

\end{document}